**Title: Silver Alloyed Wide Bandgap (Ag,Cu)(In,Ga)S$_2$ Thin Film Solar Cells With 15.5% Efficiency**

*Authors: Yucheng Hu*, Ece Washbrook, Arivazhagan Valluvar Oli, Andrea Griesi, Yurii P. Ivanov, Mariam Pelling, Simon M. Fairclough, Kulwinder Kaur, Michele Melchiorre, Adam Hultqvist, Tobias Törndahl, Wolfram Hempel, Wolfram Witte, Giorgio Divitini, Susanne Siebentritt, Rachel A. Oliver, Gunnar Kusch**

Y. Hu, E. Washbrook, M. Pelling, S. M. Fairclough, R. A. Oliver, G. Kusch

Department of Materials Science and Metallurgy, University of Cambridge, 27 Charles Babbage Road, Cambridge CB3 0FS, UK

Correspondence Emails: yh478@cam.ac.uk, gk419@cam.ac.uk

A. V. Oli, K. Kaur, M. Melchiorre, S. Siebentritt

Laboratory for Photovoltaics, Department of Physics and Materials Science Research Unit, University of Luxembourg, 41 rue du Brill, L-4422 Belvaux, Luxembourg

A. Griesi, Y. P. Ivanov, G. Divitini

Electron Spectroscopy and Nanoscopy, Istituto Italiano di Tecnologia, Via Morego 30, I-16163 Genoa, Italy

A. Hultqvist, T. Törndahl

Ångström Solar Center, Division of Solar Cell Technology, Department of Materials Science and Engineering, Uppsala University, Regementsvägen 10, SE-75121 Uppsala, Sweden.

W. Hempel, W. Witte

Zentrum für Sonnenenergie- und Wasserstoff-Forschung Baden-Württemberg (ZSW), Meitnerstraße 1, 70563 Stuttgart, Germany

Y. Hu, E. Washbrook, and A. V. Oli contributed equally to this work.

Funding: We would like to acknowledge funding from the Engineering and Physical Science Research Council (EPSRC) under EP/V029231/1 and EP/R025193/1. This work was also funded by Luxembourgish Fond National de la Recherche (FNR) for REACH (Project no: INTER/UKRI/20/15050982) and European Union within the SITA project (no. 101075626).

Keywords: Sulfide chalcopyrite, Cu(In,Ga)S$_2$, solar cells, tandem cells, photoluminescence, cathodoluminescence, electron backscatter diffraction

**Abstract**

Sulfide chalcopyrite Cu(In,Ga)S$_2$ (CIGS) is a wide bandgap semiconductor suitable for the top cell of a tandem solar device. Here we demonstrate significant improvements in absorber quality by alloying with Ag to form (Ag,Cu)(In,Ga)S$_2$ (ACIGS) absorbers. We report the Ag alloying effect on compositional, structural, and optoelectronic properties of absorbers. We demonstrate suppressed bulk recombination and improved carrier lifetime in ACIGS, as a result of improved grain size, porosity reduction and defect passivation. We also show that Ag alloying flattens the Ga gradient. Consideration of this impact of Ag will be necessary in future engineering of the Ga profile to maximize charge carrier collection and avoid interface recombination. Exploiting the beneficial effects of Ag alloying, we report a wide bandgap (1.58 eV) ACIGS solar cell with a high power conversion efficiency of 15.5% and a large open-circuit voltage ($V_{OC}$) of 948 mV, improving on the reference pure CIGS solar cell, with an 11.2% efficiency and an 821 mV $V_{OC}$. Ag alloying is a useful route to further increase the efficiency of CIGS solar cells and future tandem devices.

## 1. Introduction

The power conversion efficiency (PCE) of single-junction solar cells is ultimately limited by two mechanisms: sub-bandgap energy transmission and thermalization of hot charge carriers.[1] These losses can be mitigated by employing tandem solar cell configurations, where a wide-bandgap top cell and a narrow-bandgap bottom cell are stacked to maximize the absorption spectrum.[2–4] The chalcopyrite Cu(In,Ga)S$_2$ (CIGS) is one of the most promising candidates for the wide bandgap, $E_g$ = 1.4 eV – 2.4 eV[5], absorber in tandem solar cells due to its environmental stability and scalability, which leverages established manufacturing technologies.[6–8] Despite its potential, the certified PCE of CIGS thin film solar cells so far remains at 15.5%[9], which is significantly lower than its selenide counterpart, Cu(In,Ga)Se$_2$ (CIGSe), which has recently reached a world record PCE of 23.63%[10], obtained by Ag alloying. Achieving further advancements in CIGS requires a deep understanding of performance-limiting factors, particularly those affecting the open-circuit voltage ($V_{OC}$) deficit.[5,11,12] The $V_{OC}$ deficit primarily arises from defects within the bulk of the absorber and, secondarily, from interface recombination due to band misalignment between absorber and subsequent buffer layer. While interface recombination can be effectively minimized by optimizing buffer layers, addressing bulk recombination necessitates a deeper understanding and mitigation strategies through compositional tuning and growth processes optimization. Engineering bulk CIGS through compositional control, bandgap grading, and grain size enhancement has shown promise in reducing the quasi-fermi level splitting ($\Delta E_F$) deficit, which reflects the non-radiative losses of the absorber.[5,11–14] High-temperature growth and Cu-poor stoichiometry have been identified as beneficial for improving the CIGS absorber's quality. For example, CIGS grown at high temperatures exhibits high $V_{OC}$, leading to higher PCE, attributed to enhanced bulk quality and improved band alignment at the buffer layer interface.[14,15] Systematic compositional engineering in graded bandgap CIGS has effectively reduced bulk recombination losses associated with deep and shallow defects, leading to a certified $V_{OC}$ of 981 mV and an active area PCE of 16.1%.[12] Controlling the Ga/(Ga+In) (GGI) ratio and its profile throughout the film thickness has also been shown to enhance the quality of the bulk region of CIGS films.[12,16]

In addition to compositional tuning, alloying with Ag in selenide chalcopyrites has demonstrated increases in grain size, carrier lifetime, and structural defect reduction, which collectively enhance solar cell performance.[10,17,18] Ag addition lowers the alloy's melting temperature, promoting better elemental interdiffusion and reducing structural and electronic disorder.[10,19] Furthermore, it increases the bandgap of CIGSe, with a rather strong bowing, such that the bandgap remains unchanged or slightly decreases at low Ag content but strongly increases at medium and high Ag content.[20–22] A comparative microstructural characterization of Ag alloyed CIGSe reveals enhanced grain size while retaining the texture characteristics of reference CIGSe.[17] Moreover, Ag alloying facilitates low-temperature growth, which enables the use of lightweight flexible substrates.[23] Recently, a high concentration Ag alloyed CIGSe solar cell with a "hockey-stick"-like Ga profile achieved the record efficiency of inorganic thin film solar cells.[10] This formulation minimizes the bandgap variation with depth, which results in reduced $V_{OC}$ losses.[24] Despite the successes observed in selenide chalcopyrites, Ag alloying in sulfide chalcopyrite based solar cells has yielded less favorable results. Early studies on Ag addition to Cu-rich $CuInS_2$ indicated minimal impact on grain size and PL emissions, with Ag being distributed inhomogeneously throughout the absorber's depth, which resulted in worse solar cell performance compared to Ag free $CuInS_2$.[25] However, later studies demonstrated an increase in grain size due to improved Ag diffusion and mobility within the grains, by in-situ monitoring the Ag assisted recrystallisation annealing process of Cu-poor $CuInS_2$ film.[26] In a recent work, Ag was introduced into CIGS via a precursor layer method, resulting in enhanced grain size and lower Urbach energy at an optimal Ag/(Ag+Cu) (AAC) ratio of 0.04.[27] Nevertheless, these improvements might be offset by the formation of voids and poor uniformity, leading to reduced photovoltaic performance compared to an Ag free absorber.[28]

The contrast in outcomes observed between selenide and sulfide chalcopyrites upon Ag addition highlights the necessity for a systematic investigation of Ag alloyed CIGS (ACIGS). Herein, we employ a 3-stage co-evaporation process to grow ACIGS via a precursor layer method. Ag precursor layers with different thicknesses of 5 nm, 10 nm, and 20 nm were used to grow ACIGS absorbers in the same batch together with a reference Ag free CIGS absorber. The Ag of the precursor layer diffuses into the co-evaporated CIGS during the growth process, producing ACIGS absorbers with differing Ag concentration while maintaining similar CIGS composition ratio across all samples. No Ag layer remains after the absorber process at elevated temperatures. We reveal the potential impact of Ag alloying on CIGS absorbers from three major perspectives: the chemical composition, the microstructure, and the optoelectronic properties. We explore the change in compositional profile, especially the GGI gradient across the depth and the notch position, as a function of the addition of Ag. In terms of microstructure, we investigate the variation of crystallinity at the nanoscale, including the grain structure and void density in absorbers with increasing Ag content. By using luminescence techniques, we study the radiative recombination behavior of absorbers, analyzing the effect of Ag alloying on minority carrier lifetime and doping density. The cathodoluminescence (CL) study also directly visualizes the variation of local radiative emission across the depth of absorber, revealing the improved lateral bandgap homogeneity and suppressed defect emission with Ag alloying. Finally, we report a champion ACIGS thin film solar cell that achieves an active area PCE exceeding 15.5%, highlighting its potential for use in tandem photovoltaics.

## 2. Results and Discussions
### 2.1. Composition and Microstructure

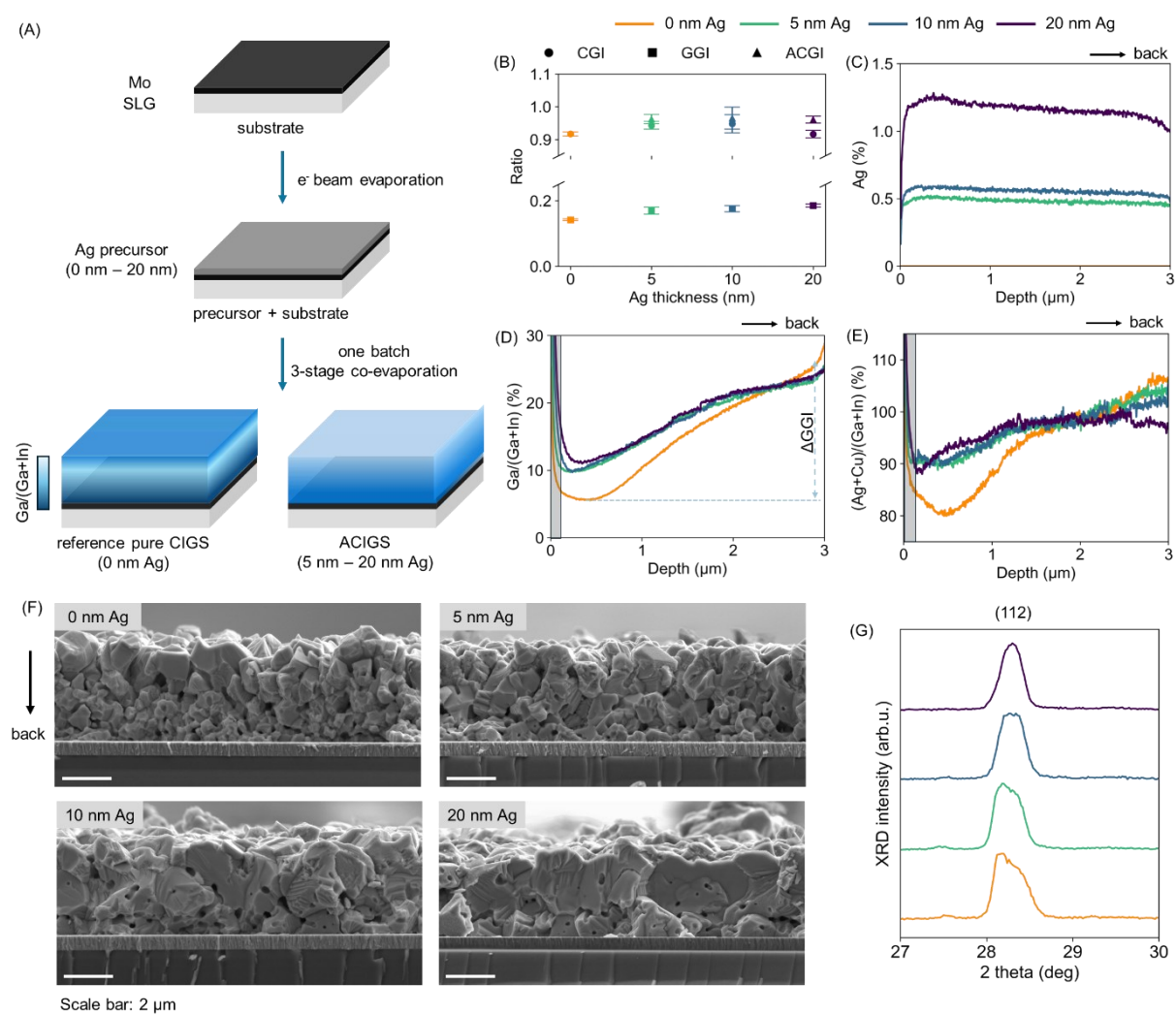

*Figure 1. Absorber fabrication and elemental analysis. (A) Schematic of CIGS and ACIGS absorbers fabrication, including the Ag precursor layer alloyed into absorbers during co-evaporation. (B) Elemental composition ratios of CIGS and ACIGS with varying Ag precursor layer thickness, as determined by EDS. (C-E) Depth profile of (C) Ag distribution, (D) bulk Ga/(Ga+In) ratio and (E) (Ag+Cu)/(Ga+In) ratio, as measured by GDOES. (F) Cleaving cross-sectional SE images of absorbers with 0 nm, 5 nm, 10 nm, and 20 nm Ag*

*precursor layer. Absorber layer is at the top; the Mo layer and soda-lime glass substrate are at the bottom. (G) XRD pattern of CIGS absorbers with varying thickness of Ag precursor layer.*

The CIGS and ACIGS absorbers were grown in the same batch with different Ag precursor layer thicknesses in a 3-stage process (deposition profile shown in Figure S1) as shown in Figure 1A. The bulk elemental composition was determined using energy dispersive X-ray spectroscopy (EDS) in a scanning electron microscopy (SEM) with an acceleration voltage of 20 kV. The atomic ratios obtained from plan view EDS for Cu/(Ga+In) (CGI), GGI, and (Ag+Cu)/(Ga+In) (ACGI) are shown in Figure 1B. The CGI and ACGI reflect the group I to group III ratio in the absorbers. There is no significant variation in GGI and ACGI ratio with increasing Ag thickness; however, the reference CIGS exhibits a slightly lower GGI ratio and both the reference CIGS and the 20 nm ACIGS show a lower CGI ratio, although all films were exposed to the same Cu, In, and Ga fluxes during the process. The lower GGI and CGI of the CIGS film without Ag are most likely due to lower Ga and Cu ratios near the surface, compared to the Ag containing films (see below). All the absorbers maintain a group I (Ag+Cu)-poor stoichiometry (ACGI < 1), which is advantageous for suppressing bulk defects and improving solar cell performance.[11,14,29] Glow-discharge optical emission spectroscopy (GDOES) was used to measure the compositional depth profile of the absorber. Figure 1C and Figure S2 present the depth profiles of Ag and AAC ratio. The Ag content in the ACIGS absorber corresponds to AAC ratios of 2%, 2.5%, and 5% for 5 nm, 10 nm, and 20 nm Ag, respectively. Ag is distributed almost evenly throughout the depth of the absorbers. Notably, the Ag and AAC profiles of the 5 nm ACIGS are very similar to the 10 nm ACIGS, suggesting a possible deviation in Ag precursor thickness as measured using a quartz crystal microbalance during e-beam evaporation of Ag. Consequently, it is estimated that the 5 nm Ag thickness may have a 2 nm to 3 nm error range. GGI and ACGI depth profiles from GDOES are shown in Figure 1D and 1E. Both GGI and ACGI are influenced by Ag alloying. The high GGI observed at the surface (marked with grey box in Figure 1D and 1E) is likely due to artefacts. The reference CIGS exhibits a broad notch, a minimum GGI point which represents $E_g$ in a graded absorber, and a steep GGI grading toward the back contact, resulting in a high variation in the GGI ratio. Ag alloying flattens the GGI grading (i.e. decrease in ΔGGI), shifts the notch upward, and moves it closer to the surface. These observations are consistent with earlier studies on CIGSe, which reported a decrease in GGI grading with Ag alloying.[19,23] The GGI profile of the Ag containing films shows a rather narrow notch very close to the surface. That is not an ideal profile for best device performance.[24] Further process optimization will have to improve this profile towards a much wider notch. The lower ΔGGI towards the back on the other hand is not an issue: a similar profile has led to a remarkably low non-radiative voltage loss.[12] The ACGI profile of most films is rather flat, as expected. Only the CIGS without Ag shows a decreased Cu content near the surface. It is possible that a Cu-poor phase formed near the surface of this film. The reason for this behavior is unclear. The profiles also offer explanations for the observations from EDS. Although 20 keV EDS examines the bulk of the films, the composition near the surface is more strongly weighted. The low Ga content and Cu ratio near the surface may explain the measured lower CGI and GGI in the CIGS film without Ag. The lower CGI in the 20 nm ACIGS film may be related to the somewhat higher Ag near the surface. Figure S3 presents the S, Cu, In, Ga, and Ag elemental depth profiles for all absorbers. It is evident that the reference CIGS is relatively In-rich near the front surface. S and Cu are almost uniformly distributed across the depth of the absorbers.

Figure 1F shows secondary electron (SE) images of cleaved cross sections of CIGS absorber layers with different thickness of Ag precursor layer, ranging from 0 nm to 20 nm. The cross sections do not cleave cleanly due to the polycrystalline nature of the CIGS absorber, but all four absorbers appear to present a grain size variation through the thickness of the film, with larger grains on the frontside and smaller grains on the backside. This observation is typical for chalcopyrite films prepared by the 3-stage process[30] and likely reflects the GGI grading in the absorber.[31] Another common feature that can be found in all four samples is the presence of voids with a non-uniform shape which appear randomly distributed. The analysis of void density will be detailed later in the context of studies on beveled cross-sections. By comparing the morphology between absorbers, we find that the overall grain size gradually increases when introducing an increasing amount of Ag to the absorber, matching literature reports for CIGSe.[21,32,33]

Figure 1G displays the X-ray diffraction (XRD) patterns of the CIGS and ACIGS absorbers, focusing on the dominant 112 reflection. The wide angle XRD pattern is shown in Figure S4, where the peaks correspond to the tetragonal chalcopyrite structure of CIGS (ICDD:00-056-1309)[14]. While the FWHM values derived from the Gaussian fits (CIGS: 0.39° to 20 nm ACIGS: 0.28°) are included for reference in Figure S4, they primarily reflect the compositional grading effects rather than microstructural characteristics like grain size or crystallinity. Therefore, we rely on SEM images of beveled films for a more accurate assessment of grain size and crystallinity. The larger grain sizes as a result of flatten grading observed in SE images correlate well with the enhanced compositional uniformity inferred from the XRD.

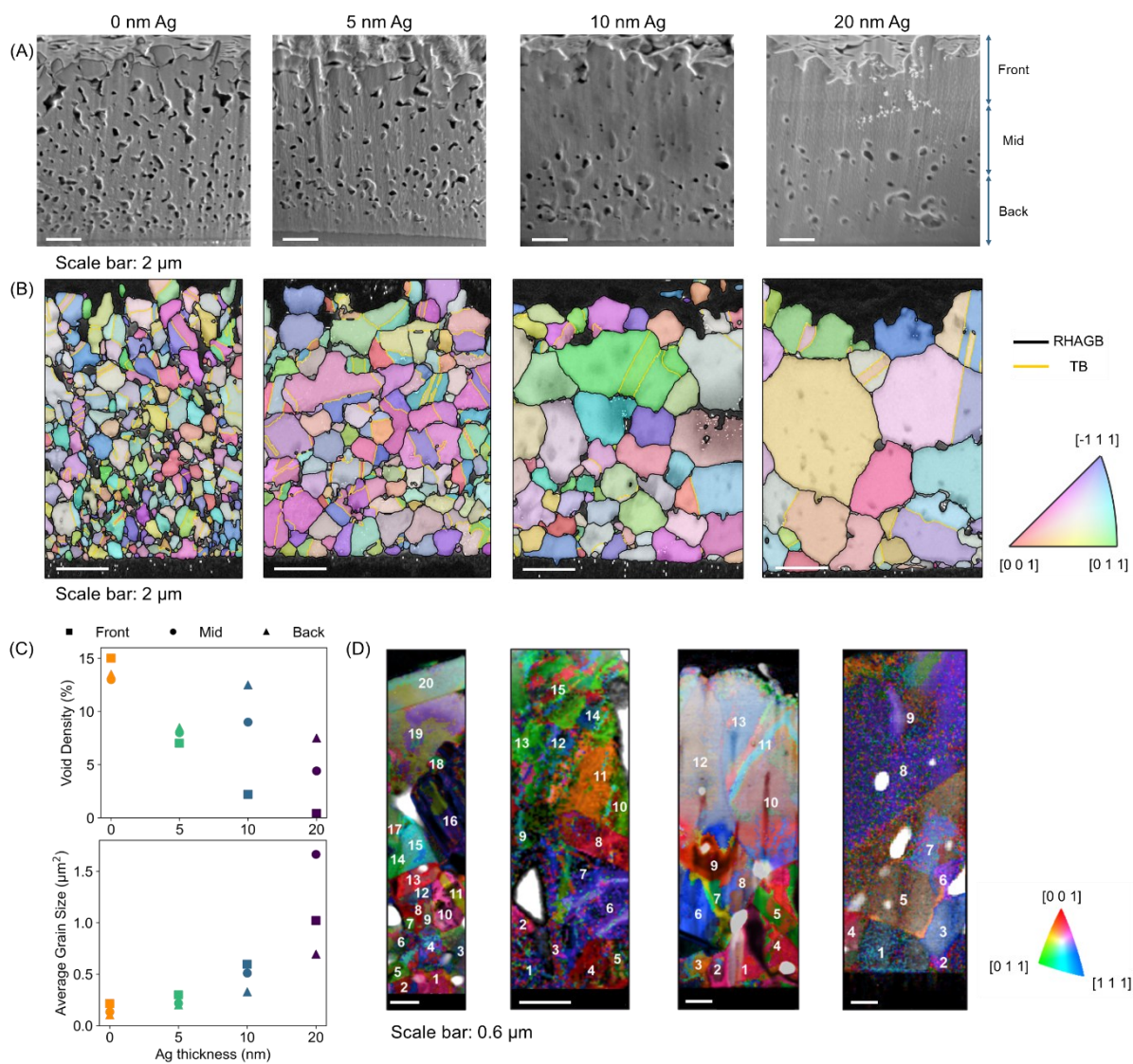

*Figure 2. Microstructure and grain structure on polished cross sections. (A) Top-view SE images of bevel surfaces of CIGS and ACIGS absorbers. (B) EBSD maps composed of grain orientation maps and band contrast maps. The orientation is color-coded in inverse pole figure Z direction. The EBSD map was 70° tilt corrected. (C) Computed average grain size and void density at different section of the SEM/EBSD maps. (D) 4D-STEM maps acquired on FIB lamellae for each sample. The image contains an overlay of the virtual bright field extracted from the dataset with the color-coded direction from Automated Crystal Orientation Mapping. White/black pixels indicate diffraction patterns that could not be indexed reliably. The bottom to up white numbers correspond to single grains.*

In order to obtain more precise cross-sectional information, we prepared cross sections using a bevel approach, as presented in Figure S5. We present a comparison of microscopic results between bevel and polished cross section in the supplementary information section 2 to highlight the benefits of the bevel approach electron backscatter diffraction (EBSD) and CL measurements. Figure 2A shows the SE images of the bevel surfaces, with clearly visible voids in the absorber layer. We observe a homogeneous distribution of voids in the pure CIGS sample and the 5 nm ACIGS sample, while the 10 nm and 20 nm ACIGS absorbers voids are primarily concentrated at the backside of the absorber. The computed void density, refers to the void area over total surface area, at each section is shown in Figure 2C. With increasing Ag content in the absorbers, the overall void density gradually decreases from 13.5% to 5%. The reduction in overall void density is mainly due to differences at the frontside of the absorber, as all absorbers exhibit comparable densities of void at the backside.

Figure 2B shows EBSD maps of the beveled cross-sections, in which the grain orientation maps are overlaid on top of the band contrast maps. The grain boundaries (GBs) are labelled as random high angel grain boundaries (RHAGBs) and twin boundaries (TBs), according to the microstructure difference. The EBSD maps provide a more reliable analysis of grain structure in absorbers than SE images of cleaved cross sections. The quantified grain size total GB lengths of the absorbers are shown in Table S2. The grain size trend extracted from the EBSD maps is generally consistent with the observations on cleaved cross-sections: the average grain size of an absorber increases as Ag content increases. Additionally, the grain size generally becomes smaller, and the total length of GBs increases from the frontside to the backside of an absorber, due to the increasing GGI ratio through the thickness of the film.[19,34] However, the computed grain size in Figure 2C shows that the region with the largest grains and the shortest GB lengths in the 10 nm and 20 nm ACIGS samples are found in the upper-middle sections, instead of directly at the surface. This valley-shaped grain size variation may be associated with the position of the notch right underneath the surface, given that sudden local compositional variation may introduce a strain to the lattice and thus limit the formation of large grains.[35,36] To exclude the potential influence of bevel geometry on grain size measurements and increase the validity of the microstructural study, we also acquired 4-dimensional scanning transmission electron microscopy (4D-STEM) maps on FIB lamellae. The 4D-STEM maps, as shown in Figure 2D, further validate the grain size enhancement induced by Ag addition and the overall reduction of grain size through the thickness of the absorber. The distinct grains have been color-coded according to their orientation and numbered in the 4D-STEM map using the python library py4DSTEM[37]; grain 11 in the 10 nm Ag absorber and grain 9 in the 20 nm Ag absorber represent examples of the relatively small grains near the absorber surface seen in the EBSD data. The observed increase in grain size can effectively reduce bulk recombination and enhance charge carrier transport ability, hence benefiting the $V_{OC}$ and short-circuit current density ($J_{SC}$) of final devices.[27,38]

The EBSD and 4D-STEM maps provide additional information about the distribution of voids within the absorbers. Voids constitute an abrupt change in local topography and hence may deteriorate local diffraction patterns in EBSD and are thus visible in band contrast maps; in 4D-STEM maps they are also visible as blank pixels. Most voids are found adjacent to GBs, which aligns with the findings in several previous reports about CIGSe.[39,40] The formation of voids is strongly associated with the agglomeration of vacancies, such as $V_{Cu}$, during the recrystallisation process of absorber growth, especially in Cu-poor absorbers.[39,40] In Ag free CIGSe absorbers grown by 3-stage co-evaporation, Cu often segregates in the form of Cu selenides into GBs and to the absorber surface at the end of Cu-rich second stage of growth.[19,40,41] During the recrystallization process, the low formation energy of $V_{Cu}$[11] and the high mobility of Cu lead to the rapid diffusion of Cu from Cu-rich areas into III-rich areas, leaving vacancies behind.[39] These vacancies may eventually aggregate and form voids. The scenario is slightly different in ACIGS absorbers: Ag alloying introduces $Ag^+$ ions, which can have low bond dissociation energy with S and may preferentially segregate at GBs.[19] Therefore, alloying with Ag lowers the melting temperature of the absorber and promotes the interdiffusion of In, Ga, and Cu during the absorber growth, hindering Cu clustering.[19,33] The formation of vacancies can be suppressed with Ag alloying.[42] The improved elemental diffusion and the reduction of vacancies ultimately leads to the reduction of voids in ACIGS absorbers. The reduction of void density should be beneficial to the performance of solar cell, since the presence of voids may lead to extra interface recombination and hence be detrimental to the $V_{OC}$ of solar cell or module.[40,43] Additionally, the voids may impair light absorption and therefore lower the $J_{SC}$ of solar cells.

## 2.2. Optoelectronic Properties

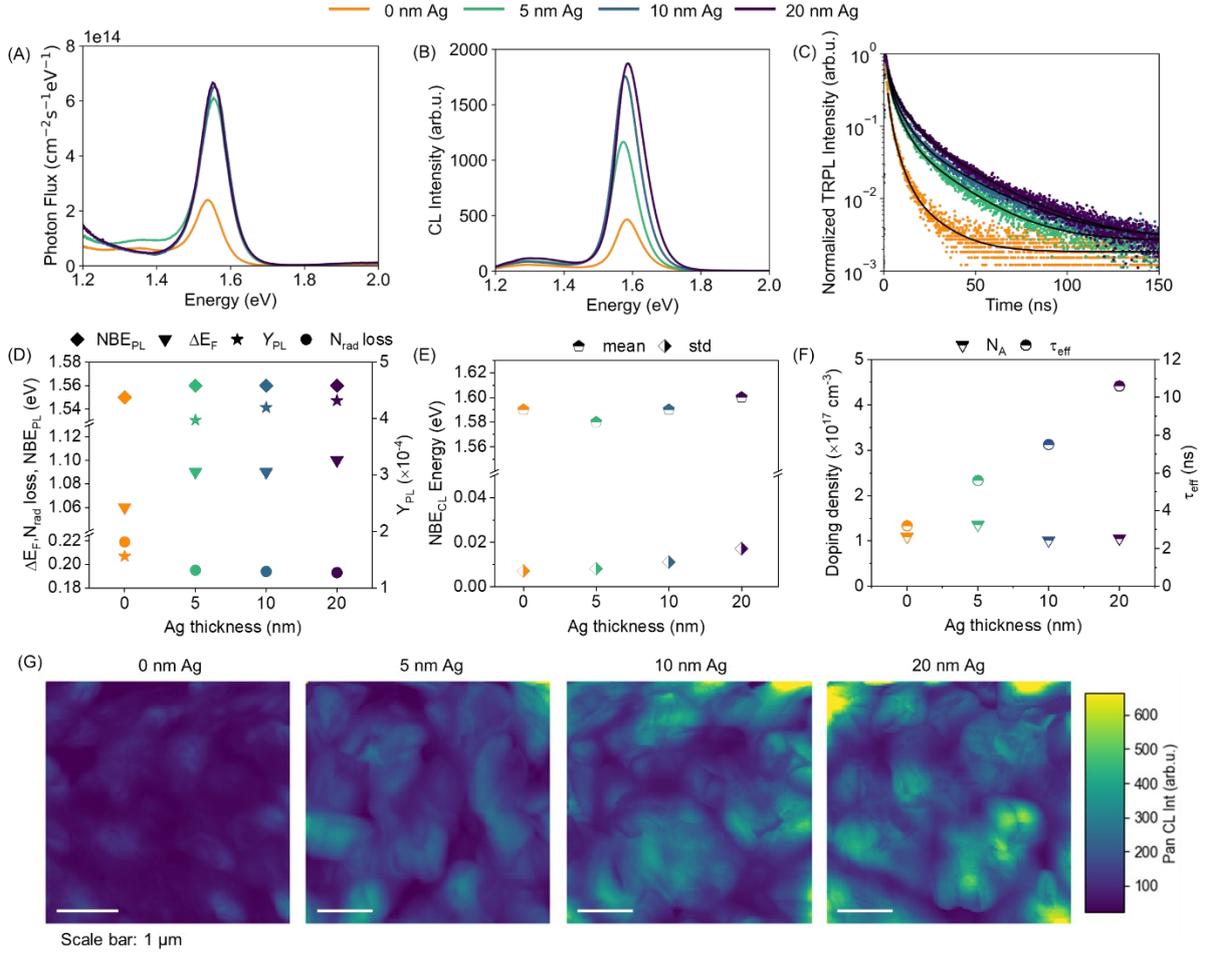

*Figure 3. Surface optoelectronic properties by PL, plan-view CL and TRPL. (A – C) (A) Absolute PL spectra, (B) mean CL spectra, and (C) TRPL transients of CIGS and ACIGS with varying Ag precursor layer thickness, the dark lines refer to the TRPL fitting curves. (D – F) (D) Extracted optoelectronic parameters from PL, (E) fitted CL NBE emission energy and standard deviation, (F) effective minority carrier lifetime and doping density for the different Ag thickness. (G) Panchromatic plan view CL intensity maps of absorbers with varying Ag thickness.*

To investigate the optoelectronic properties of absorbers, we performed both PL and CL measurements. PL is an efficient technique for quickly assessing the quality of an absorber material and identifying bulk defects prior to the complete fabrication of a solar cell. Moreover, absolute PL enables the prediction of maximum $V_{OC}$ achievable in a finalized solar cell.[44,45] The absolute PL spectra of the absorbers, illuminated with a photon flux equivalent to 1 sun, are presented in Figure 3A. The corresponding optoelectronic parameters derived from the PL, such as $NBE_{PL}$, $\Delta E_F$, and $N_{rad}$loss (which refers to Shockley-Queisser (SQ)-$V_{OC}$-$\Delta E_F$), are shown in Figure 3D. The PL spectrum of our Ag free CIGS exhibits a PL maximum peak (referred to as $NBE_{PL}$) at ~1.55 eV, corresponding to near band edge (NBE) emission, along with a shallow defect emission (DE) at ~1.3 eV.[11,46] For 5 nm ACIGS, the NBE$_{PL}$ intensity increases, accompanied by a ~10 meV blueshift, and the DE is suppressed. Further increasing Ag thickness to 10 nm and 20 nm seems to have only a small effect on the NBE intensity, while the DE decreases. It has been proposed that substitution of Cu with Ag in chalcopyrite and kesterite materials can lead to changes in both the valence band maximum and the conduction band minimum, with the exact changes being strongly dependent on the specific material system and the alloying concentration, ultimately resulting in an increase in the $E_g$.[22,27,47–49] In our study, Ag alloying results in an decreased notch depth of GGI profile compared to CIGS (Figure 1D) but has minimal impact on the measured bandgap energy NBE$_{PL}$, consistent with the minimal change in $E_g$ observed in chalcopyrites when adding relatively small amounts of Ag or Ga to CuInS(e)$_2$.[25,33,50] However, across all ACIGS absorbers, the trend between notch depth and NBE$_{PL}$ remains consistent, exhibiting similar $E_g$ corresponding to notch depth. We anticipate that further increasing the AAC ratio while maintaining (Ag+Cu)/(In+Ga)-poor stoichiometry could attribute to an increase in the $E_g$,[18,21,51] leading to potential better matching with Si or CIGSe low bandgap bottom solar cell in tandem configuration.[3,4] The PL quantum yield ($Y_{PL}$) method was used to determine the $\Delta E_F$.[44,45] The $Y_{PL}$ was calculated using the relation, $Y_{PL}=\Phi_{PL}/\Phi_{laser}$, where $\Phi_{PL}$ represents the integrated flux of the PL peak within the 1.4 eV to 2.0 eV range, and $\Phi_{laser}$ is an incident photon flux corresponding to the number of photons in a 1 sun spectrum above the $E_g$. Using the laser flux instead of the absorbed laser flux, underestimates the $Y_{PL}$ and thus $\Delta E_F$ by about 5 meV, assuming a reflectance of 10% to 20%. The calculated $Y_{PL}$ for CIGS and ACIGS are presented in Figure 3D. An increase in $Y_{PL}$ indicates that non-radiative losses are reduced with Ag alloying. Subsequently, $\Delta E_F$ was determined using the relationship of the SQ limit and non-radiative recombination, as described in reference[1] and expressed in the Equation 1,

$$\Delta E_F = qV_{OC}^{SQ} + k_B T \times \ln Y_{PL} \qquad \text{(Equation 1)}$$

where $qV_{OC}^{SQ}$ is the SQ-$V_{OC}$ limit, $k_B$ is the Boltzmann constant, and $T$ is the temperature. The $E_g$ for $qV_{OC}^{SQ}$ is taken from the PL maximum, however, we note that this approach further underestimates the $\Delta E_F$ by typically 10 meV to 20 meV.[44] As anticipated, both the $\Delta E_F$ and the non-radiative losses are constant across all ACIGS absorbers, which aligns with the similar PL peak positions and intensities observed. The calculated $\Delta E_F$ for 5 nm, 10 nm and 20 nm ACIGS are 1.09 eV, 1.09 eV and 1.1 eV, respectively, which is over 40 meV higher than that of CIGS, suggesting the potential for achieving a higher $V_{OC}$ with a suitably band aligned buffer layer.[44,45] Furthermore, the non-radiative loss decreases from 219 meV (CIGS) to ~194 meV for all the ACIGS, indicating a reduction in deep defects due to Ag alloying, which reduces structural disorder.[18]

CL is a SEM-based luminescence technique allowing microscopic investigation of recombination activities and defects. We performed plan view CL measurements on four absorbers using the same measurement conditions for all data sets. Similar to PL results, mean CL spectra and panchromatic CL maps (Figure 3B and 3G) show an enhanced emission intensity with increasing Ag precursor thickness. However, the mean CL spectra (Figure 3B) exhibit a variation in NBE emission energy and increasing NBE FWHM with increasing Ag content; the detailed FWHM parameter of the mean spectra are provided in Table S4. The differences between CL and PL spectra are likely due to the different excitation intensity.[52] To quantitatively analyze the dependence of the emission homogeneity on Ag content, we computed the standard deviation of the NBE emission energy in each pixel for mapped samples. We found that the standard deviations of the NBE emission energy for CIGS and 5 nm ACIGS are about 8 meV, smaller than 11 meV for 10 nm ACIGS and 17 meV for 20 nm ACIGS (Figure 3E). We attribute the increased lateral inhomogeneity with enriched Ag content to two possible factors: local compositional variation and the large variation in surface topography for the 10 nm and 20 nm sample. The rough surface, as confirmed by plan view SEM images in Figure S6, may strongly influence the probing depth of CL, leading to variations in signal due to the through-thickness compositional gradients present in all samples.[46] We therefore carried out further CL measurements on polished bevel surfaces, shown in Figure 4, which show that all the samples have roughly similar standard deviations of the NBE emission energy at the frontside (quantified data shown in Table S5), confirming that the data recorded on the rough, unpolished surface are influenced by the surface morphology rather than in-plane inhomogeneity.

Increased $\Delta E_F$ and thus higher PL and CL emission intensity can be due to longer minority carrier lifetimes and/or due to increased doping density.[44] Time-resolved PL (TRPL) gives an indication of changes in carrier lifetime. TRPL transients of the absorbers were measured at the lowest excitation intensity possible to ensure that we predominately measure minority carrier lifetimes and are shown in Figure 3C. From the transients it is already obvious that Ag addition increases the minority carrier lifetime. The decays are not exponential, therefore we use a tri-exponential fit. The effective minority carrier lifetime $\tau_{eff}$ is obtained from the fit using the weighted average method[53],

$$\tau_{eff} = \frac{A_1\tau_1 + A_2\tau_2 + A_3\tau_3}{A_1 + A_2 + A_3} \qquad \text{(Equation 2)}$$

where, the $A_i$ is the pre-factor corresponding to the lifetime $\tau_i$. The $\tau_{eff}$ values for the absorbers are presented in Figure 3F, with the corresponding fitted values summarized in Table S3. The fitted lifetimes $\tau_1$ $\tau_2$, and $\tau_3$ of all ACIGS are higher than reference CIGS. The increase in all fitted lifetimes for ACIGS is attributed to an improved morphology with larger grains, reduced grain boundaries, and reduced non-radiative losses.[54,55] The $\tau_{eff}$ values calculated from Equation 2 are 3.5 ns, 5.6 ns, 7.5 ns and 10.6 ns for CIGS, 5 nm ACIGS, 10 nm ACIGS and 20 nm ACIGS, respectively. The long $\tau_{eff}$ or $\tau_3$ (particularly for 10 and 20 nm ACIGS) is among the highest in the literature for CIGS[11,13,56], which aligns with the observations that the ACIGS has a larger grain size of approximately 1 µm and also with the XRD data that suggest improved crystallinity (see Figure 1 and Figure 2). A high minority carrier lifetime is crucial for enhancing charge carrier generation and extraction, leading to an increased $V_{OC}$ and $J_{SC}$.[57,58]

To investigate if the increase in lifetime is sufficient to explain the increased quasi-Fermi level splitting, we extract the doping density, $N_A$, of the absorber, by relating $\Delta E_F$ to $\tau_{eff}$ [13,59],

$$N_A = p_0 = \frac{d \cdot N_C N_v}{G \cdot \tau_{eff}} exp\left(\frac{\Delta E_F - E_g}{k_B T}\right) \quad \text{(Equation 3)}$$

where $N_c$ and $N_v$ represent the effective density of the states of the conduction and valence bands, respectively, for which we have used the values corresponding to $CuInS_2$[15]. $G$ is the generation flux during steady state $\Delta E_F$ measurement and $d$ denotes the thickness of the films (2.8 µm). The calculated $N_A$ of the absorbers are presented in Table S3. Doping density is reported to decrease with Ag alloying in selenide chalcopyrites.[18,51] In our study, the doping density does not change upon Ag alloying, and ACIGS absorbers still maintain a high $N_A$ level with even 2.5% Ag incorporation (ACGI remains <1), which can be beneficial for achieving good $V_{OC}$ of ACIGS devices. The doping level of all absorbers is rather high, which could be due to the rather high ACGI ratio at a low Ag content.

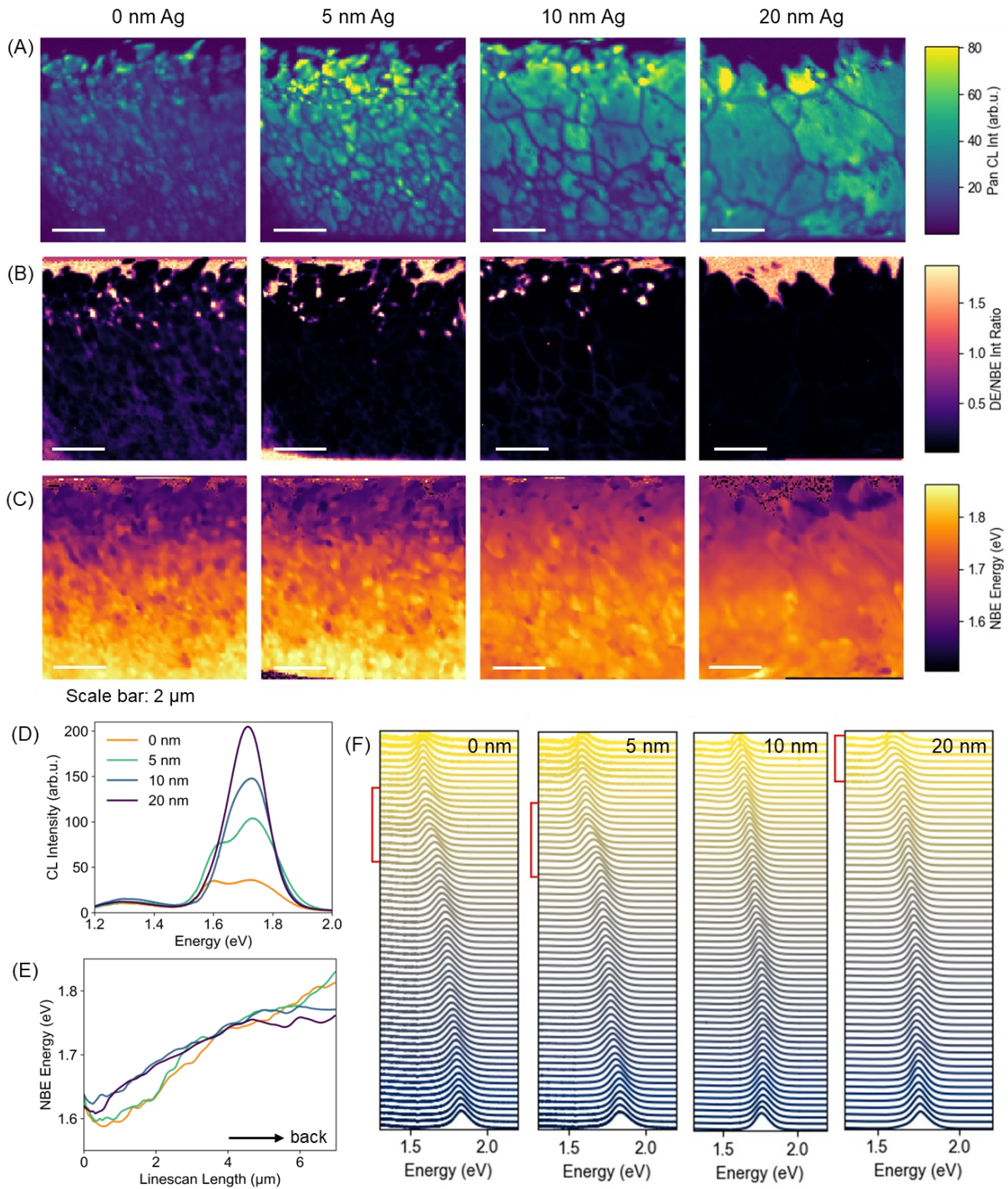

*Figure 4. Micro-scale opto-electronic properties by CL. (A – D) (A) Panchromatic CL maps, (B) defect intensity to NBE intensity ratio maps, (C) NBE emission energy maps, and mean CL spectra (D) for CIGS and 5 nm, 10 nm, and 20 nm ACIGS absorbers. (E) Averaged line profile of the NBE emission energy variation. (F) Extracted CL spectra along the depth of the bevel, the highlighted regions indicate band broadening around the notch region.*

To further investigate the recombination pathways through the thickness of the absorber and study the variation of optoelectronic properties at a microscopic scale, we acquired CL maps on the beveled cross sections in the same area as the EBSD maps. The variation in aspect sizes between CL and EBSD maps is mainly due to the difference in electron beam incident direction, as shown in Figure S5. The bevel CL maps are not affected by the topography of the original film surface, allowing the impact of GBs on light emission to be fully appreciated. Other factors, such as light extraction efficiency variations, can affect the interpretation of hyperspectral CL data from unpolished, rough absorbers. However, in terms of comparison between PL and CL results, it is worthwhile to note that the plan view data in Figure 3B and 3G are more directly comparable with PL since they sample the near surface region more extensively. Figure 4A shows the panchromatic CL maps of the four bevels, revealing the variation of the CL emission intensity between different absorbers. In Figure 4B and 4C we plot the results of fitting Gaussian functions to each pixel in the hyperspectral maps, extracting the distribution of the DE to NBE peak intensity ratio ($I_{DE}/I_{NBE}$) as well as the NBE peak emission energy. By comparing the CL results with the corresponding EBSD maps (Figure 2), the dark boundaries in the panchromatic CL map (Figure 4A) can be confirmed as RHAGBs. These RHAGBs appear to be bright in DE/NBE Intensity map, suggesting strong DE at RHAGB positions. With the increased grain size and reduced GBs length, the ACIGS absorbers generally exhibit higher panchromatic CL intensity. In addition, the RHAGBs in the ACIGS absorbers generally exhibit a lower DE/NBE intensity ratio, indicating the reduction of deep defect recombination in ACIGS absorbers, in agreement with PL and plan view CL measurements (Figure 3). The observation of reduced defect recombination with Ag alloying in CL aligns with the improved $Y_{PL}$ derived from PL. The inhibited defect recombination in ACIGS can be attributed to a reduced density of deep defects and/or passivation of deep defects.[18] TBs, as identified by the EBSD maps, do not affect the local optoelectronic properties strongly and thus are not prominent in CL maps. The observation has been noted in previous chalcopyrite research and is related to the high symmetry structure of TBs.[60–62] It is also noted that no high bandgap phase, like an ordered defect compound (ODC) is observed at the surface of any of the films within the resolution of the applied methods like CL.
Apart from GBs, a considerable amount of high DE spots located close to or on the GBs can be found near the notch depth in all bevels except the 20 nm ACIGS sample. The absence of these spots in 20 nm ACIGS is likely to be a local variation because a few high DE spots were observed in the 20 nm ACIGS when a cleaved cross section was examined in CL, as presented in Figure S7. However, the combination of bevel and cleaved cross section CL results still indicates that the density of spots with high DE intensity is greatly reduced in high Ag content absorbers. A potential explanation for the formation of these spots of high DE intensity could be the aggregation of defects induced by the relatively high lattice strain around the notch positions.[35] As shown by the GDOES results (Figure 1D and 1E), the addition of Ag flattens the compositional variation around the notch position, which could lead to potential relaxation of lattice distortion and strain[22,63], hence reducing the formation of high DE spots.

The NBE energy maps in Figure 4C provide straightforward visualizations of $E_g$ variation across the absorbers. All absorbers show an obvious increase in the $E_g$ from the top to the back, but the grading in Ag free CIGS and 5 nm ACIGS absorbers is more uneven than for ACIGS absorbers with higher Ag content. The computed standard deviation of the NBE energy quantified such a spatial change in bandgap grading at micro-scale. The NBE energy standard deviation in the CIGS and 5 nm ACIGS are 81 meV and 80 meV, which are much higher than that for the 10 nm and 20 nm ACIGS absorbers at 51 meV and 52 meV, respectively. The reduced NBE peak energy variation can be attributed to Ag alloying induced improvements in elemental diffusivity and hence homogeneity.[19,63] The variation in standard deviation of the NBE energy is in alignment with the peak shape and FWHM change of the NBE peak in mean CL spectra (Figure 4D). CIGS and 5 nm ACIGS absorbers mean spectra have obvious double peaks with large FHWM at the NBE emission region, whereas 10 nm and 20 nm only show one NBE peak with relatively small FWHM. The double peak feature indicates the coexistence of strongly distinct Ga-poor and Ga-rich regions, suggesting large elemental inhomogeneities in CIGS and 5 nm ACIGS absorbers, in agreement with the XRD (Figure 1). The NBE double peak feature appears in mean spectra because of averaging spectra over area with different composition, and only one single NBE peak observed in local CL spot spectra.

Due to local inhomogeneity, the notch positions are hard to directly pick out from the NBE energy maps (Figure 4C). Therefore, we extracted the NBE energy line profiles across the absorber, as shown in Figure 4E. The line profiles are integrated over most of the maps in lateral direction, as indicated in Figure S8. The notch positions can be found in the first 1 μm of the line profiles. The actual notch positions may be slightly deeper than the positions deduced from the line profiles as the ~150 nm interaction volume of this CL experiment also probes material beneath the bevel surface. Compared with CIGS and 5 nm ACIGS absorbers, the energy line profile for 10 nm ACIGS absorber shows significantly flattened bandgap grading with upward-shifted notch energy and shallower notch position, which is generally consistent with the compositional trend shown in GDOES GGI ratio. The flattened bandgap grading can mitigate lattice distortion and reduce the density of bulk defects.[63,64] However, these profiles are far from ideal and need optimization in further studies. 20 nm ACIGS exhibits similar changes to notch energy profiles and depths, but with the flattest backside grading. Comparing the line spectra with the GDOES results (Figure 1), the extra flat back grading in 20 nm ACIGS may be associated with the lower ACGI ratio near the back of this sample. Poor back grading can lead to increased backside recombination and thus a loss in $V_{OC}$.[59,65] It is worth noting that the disagreement between the CL results and GDOES profile for the 5 nm ACIGS absorber may arise from batch-to-batch differences. Figure S9 and S10 shows the STEM-EDS mapped elemental distribution of the batch for EBSD-CL measurements.

The line spectra in Figure 4F visualize the shift of normalized CL spectra through the film thickness. The changes in spectral shape can provide additional information about local and microscopic homogeneity variations, which cannot be offered by GDOES.[33,66] As highlighted by the red brackets beside the line spectra, NBE peak broadening can be found around the notch region of several absorbers. The broadening effect is obvious in CIGS and 5 nm ACIGS and is rather subtle in 20 nm ACIGS. The 10 nm ACIGS absorber shows the smoothest peak shift toward the backside. The peak broadening indicates the potential co-existence of low bandgap Ga-poor region and high bandgap Ga-rich region around the notch positions; this behavior has also been reported in a previous study[66]. The local distinct GGI change can be confirmed by comparison with STEM-EDS maps (shown in Figure S9), in which the CIGS and 5 nm ACIGS absorbers show intermixing Ga-rich and Ga-poor areas while 10 nm and 20 nm ACIGS exhibit a more uniform lateral composition. The smooth grading in 10 nm and 20 nm ACIGS absorbers, together with the flattened NBE energy profile, further confirms that the overall homogeneity of absorbers can be improved with an adequate amount of Ag alloying.

To summarize, the combination of PL, TRPL, and CL reveals the suppression of defect recombination and improved lifetime by Ag alloying, leading to better $\Delta E_F$. The optoelectronic analysis also demonstrates the influence of Ag alloying on Ga grading, which, however, may need further optimization to achieve a wider notch and better back grading.

## 2.3. Solar Cell Performance

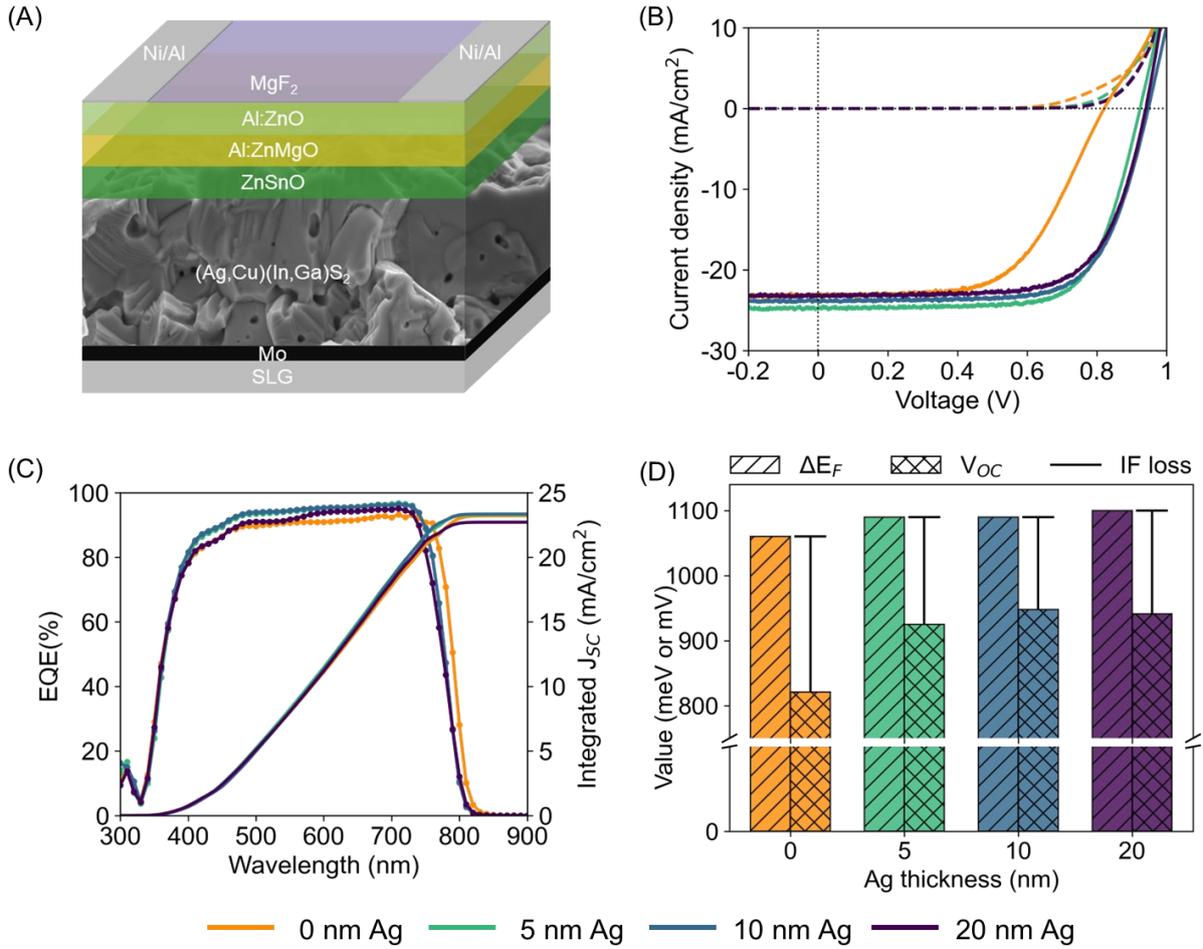

*Figure 5. Device performance of CIGS and ACIGS solar cells. (A) Schematic of the ACIGS solar cell with 10 nm ACIGS cross sectional SEM absorber. (B) J-V characteristics of CIGS and ACIGS solar cells with different Ag precursor layer thickness. The solid line refers to the J-V measured under light illumination, and the dotted line are dark J-V.(C) EQE spectra and integrated $J_{SC}$ of CIGS and ACIGS solar cells. (D) Interface $V_{OC}$ loss calculated from $\Delta E_F$ - $V_{OC}$.*

Finally, we fabricated solar cells incorporating a ZnSnO$_x$ buffer layer with a Sn/(Sn+Zn) atomic ratio of 0.19, using atomic layer deposition in combination with a sputtered Al:(Zn,Mg)O$_x$ i-layer, to avoid barriers for the forward current and fill factor (FF) losses.[67] The device configuration is given by Mo/ACIGS/ZnSnO$_x$/(Zn,Mg)O$_x$/Al:ZnO/MgF$_2$ as illustrated in Figure 5A. ACIGS was grown by co-evaporating CIGS on Ag precursor layer with different thickness as discussed previously. The top three layers (from top to bottom) function as an antireflection coating, transparent front contact, and high-resistance layer, respectively. Figure 5B presents the current-density-voltage (J-V) characteristics of the champion solar cells for each amount of Ag, while Figure 5C displays the corresponding external quantum efficiency (EQE) spectra alongside integrated $J_{SC}$. The $E_g$ extracted from the inflection point of the EQE spectra for all absorbers are presented in Table S6. The $E_g(EQE)$ for CIGS is 1.56 eV, similar to $NBE_{PL}$, whereas all ACIGS absorbers exhibit an $E_g(EQE)$ of 1.58 eV, which is over 20 meV higher than $NBE_{PL}$. The larger PL shift away from the absorption edge, i.e. the $E_g(EQE)$, may be in agreement with the increase in FWHM of the plan view mean CL peak and the increased standard deviation of the $NBE_{CL}$ distribution (although we note that some of the FWHM in CL should be attributed to the impact of surface roughness).[24,68] The photovoltaic parameters, $V_{OC}$, $J_{SC}$, FF, active area PCE, and $V_{OC}$ deficit, are detailed in **Table 1** and the interface (IF) loss ($\Delta E_F - V_{OC}$) are depicted in Figure 5D. $V_{OC}$ deficit is calculated from $SQ\text{-}V_{OC} - V_{OC}$, where $SQ\text{-}V_{OC}$ is defined based on $E_g(EQE)$. Statistical comparison of the devices is shown in Figure S11, where the box plots depict the distribution of the measured parameters. The PCE of the solar cells was calculated based on the $J_{SC}$ determined from the EQE due to the spectral mismatch of the solar simulator.[50,51] The J-V characteristics reveal that Ag alloying leads to significantly improved $V_{OC}$ and FF. The $J_{SC}$ (from EQE) of the 5 nm and 10 nm ACIGS solar cells is almost unchanged compared to the CIGS cell. However, since the bandgap is higher, this is actually an improvement. This enhanced EQE indicates improved carrier collection in the space-charge region (SCR) and increased diffusion length in the higher wavelength region.[51] In contrast, the EQE of 20 nm ACIGS decreases in the lower wavelength region. Such behavior is typically observed in weakly doped absorbers, where the SCR is large and the p=n point deep in the absorber.[69] This is not the case in these highly doped absorbers. However, in every well-functioning heterojunction solar cell the absorber surface is inverted, i.e. holes are the minority carriers near the interface.[69,70] If the interface of the 20 nm ACIGS shows a higher recombination rate for holes, this loss in short wavelength EQE might be explained. This explanation would also not contradict the higher PL and TRPL results seen in 20 nm ACIGS compared to 10 nm ACIGS, since the PL measurements are performed on absorbers film without SCR. We speculate that slightly high Ag concentration near the surface, as indicated by GDOES (see Ag and AAC depth profile shown in Figure S2 and S3), might be behind this difference. The significant IF loss observed in CIGS, i.e. $V_{OC}$ relative to $\Delta E_F$[71], is likely attributed to two main factors: (i) high concentration of surface defects extending into the SCR, as indicated by the CL defect ratio map, that increase recombination at or near the interface, (ii) shifts in the conduction band minimum compared to ACIGS, resulting in band misalignment with the buffer layer. The champion solar cell based on 10 nm ACIGS achieved an active area PCE of 15.5% with the $V_{OC}$ of 948 mV. The IF loss for 10 nm ACIGS solar is 142 mV, which is ascribed to surface defects, as was shown in CL results. We expect that reducing the surface defects further by optimizing growth condition in the third stage of ACIGS growth and improving the GGI profile similar to the one obtained in the best CIGS absorbers[12] can lead to further improved $V_{OC}$ and thus solar cell performance.

| Ag thickness (nm) | $E_g$(EQE)[a] (eV) | $V_{OC}$[b] (mV) | $J_{SC}$[c] (mA/cm$^2$) | FF[d] (%) | PCE[e] (%) | IF loss[f] (mV) | $V_{OC}$ deficit[g] (mV) |
|---|---|---|---|---|---|---|---|
| 0 | 1.56 | 821 | 23.2 | 58.9 | 11.2 | 239 | 351 |
| 5 | 1.58 | 925 | 23.3 | 70.7 | 15.3 | 165 | 266 |
| 10 | 1.58 | 948 | 23.4 | 70.1 | 15.5 | 142 | 243 |
| 20 | 1.58 | 941 | 22.7 | 69.7 | 14.9 | 159 | 250 |

*Table 1: Solar parameters of the champion devices with different Ag thickness. All parameters rounded to 3 significant figures.*

*a Bandgap derived from EQE curve*

*b Photovoltage*

*c Photocurrent*

*d Fill factor*

*e Active power conversion efficiency[12]*

*f Interface loss*

*g Photovoltage deficiency, defined as SQ-$V_{OC}$ - $V_{OC}$*

## 3. Conclusion

In summary, we reveal the beneficial effects of Ag alloying on 3-stage co-evaporated sulfide chalcopyrite absorbers. We demonstrate the influence of Ag alloying on the composition, the microstructure, and the optoelectronic properties of the resulting film. We observe a flattening of Ga grading and a shift of the notch position closer to the surface in both compositional profiles and bandgap linescans, obtaining improved elemental homogeneity through Ag incorporation. However, further improvement of the Ga profile is needed in the future. We show that Ag alloying can improve the crystallinity of absorbers, reducing porosity, enlarging grain size, and suppressing defect-rich GBs. Combining various optoelectronic analyses, we reveal the passivation of bulk defects and enhancements of carrier lifetime by Ag substitution. No change in doping level is observed. The combined effect of microstructure and defect passivation on the reduction of bulk recombination contribute to an increase in $\Delta E_F$ and hence improvements in device $V_{OC}$. We also demonstrate a reduction in microscopic cross-sectional inhomogeneity, which may enhance local recombination in ACIGS absorbers. We demonstrate a 15.5% active area PCE champion ACIGS solar cell with high $V_{OC}$ of 948 mV, a good FF of 70.1%, and high $J_{SC}$ of 23.4 mA/cm$^2$ for a bandgap of 1.58 eV.

Ag alloying to enlarge grain size and reduce bulk recombination is an effective engineering approach to consistently achieving CIGS solar cell efficiencies over 15%. In order to push PCE further, strategies such as optimizing the third stage of 3-stage growth process, and incorporating alkali treatments can further reduce the $\Delta E_F$ deficit, while buffer layer optimization minimizes IF loss and increases FF, collectively advancing the development of high efficiency CIGS absorbers for tandem solar cells.

## 4. Methods

The absorbers were grown using a 3-stage process previously reported for both sulfide and selenide chalcopyrites.[11,72] The deposition profile is illustrated in Figure S1. CIGS and Ag alloyed CIGS (ACIGS) were grown on molybdenum (Mo) coated soda-lime glass (SLG) substrates. The actual substrate temperatures ($T_S$) were ~470°C for the first stage and ~600°C for the second and third stages. $T_s$ was estimated based on the substrate heater calibration using a pyrometer and known softening temperature of glass. The ramp rate was set to 20°C/min, and the substrates were rotated at 8 rpm during deposition. For ACIGS absorbers, Ag precursor layers with different thickness of 5 nm, 10 nm, and 20 nm were deposited on the Mo surface using e-beam evaporation prior to CIGS growth.[17,55] Both bare Mo and different Ag precursor layer coated Mo substrates were loaded onto the same sample holder to ensure consistency in the growth process. Prior to evaporation, the Mo or Mo/Ag surface was sulfurized with sulfur vapor at $T_S$ of 400°C for 15 minutes. The 3-stage process is illustrated in Figure S1. In the first stage, calibrated fluxes of Ga and In were co-evaporated with sulfur partial pressures ranging from $2 \times 10^{-5}$ to $8 \times 10^{-5}$ mbar. During the second stage, a calibrated flux of Cu was evaporated. Approximately after 8.5% excess (Cu+Ag) rich composition, as monitored by increases in heating power and pyrometer readout, Ga and In were co-evaporated again (without Cu) to achieve a (Cu+Ag)-poor stoichiometry.[12] In the third stage, the Ga flux was reduced compared to the first stage. The Ag precursor layer serves as a source during growth, forming ACIGS bulk absorbers with varying Ag concentration while maintaining consistent CIGS composition across samples. Due to the limitation of chamber size, three batches of absorbers were produced for different measurements. Therefore, subtle behavior differences may be observed from batch to batch. Good consistency across batches can be confirmed by XRD. For clarity, absorber batch 1 was used for PL, TRPL, and solar cell performance measurements, absorber batch 2 was used for CL, EBSD, 4D-STEM, and STEM-EDS, whereas batch 3 was used for GDOES.

Deposition of the $ZnSnO_x$ buffer layer on ACIGS/Mo stacks for solar cell devices were performed using thermal ALD at 120 °C largely following a previously developed process.[73] The resulting Sn/(Sn+Zn) ratio of 0.19 (measured using XRF) was achieved by using a supercycle approach alternating between ZnO subcycles and $SnO_x$ subcycles in a $ZnO/ZnO/SnO_x/ZnO/SnO_x$ sequence. The subcycles used a metal precursor/$N_2$-purge/$H_2O$/$N_2$-purge sequence with pulse and purge times of 0.4/0.8/0.4/0.8 s and where the metal precursor was $Zn(C_2H_5)_2$ for ZnO and $Sn(N(CH_3)_2)_4$ for $SnO_x$. In total 1500 cycles (300 supercycles) were performed resulting in a film thickness of 50-60 nm. Al:MgZnO i-layer and Al:ZnO transparent front contact were deposited using magnetron sputtering on top of the buffer layer.[12] The device is completed with evaporated Ni/Al grid electrode followed by ~100 nm thick $MgF_2$ antireflection film. Individual cells are defined by mechanical scribing with an active area of ~0.42 $cm^2$.

The cross-sectional morphology of the absorbers was imaged using SEM and the bulk compositions were determined using EDS with an operating voltage of 7 kV and 20 kV. Structural characterization was performed using X-ray diffraction with Cu-$K_\alpha$ radiation and the data was plotted without applying instrumental resolution correction.

The compositional depth profiles were recorded by GDOES. An argon plasma at a pressure of 450 Pa erodes the sample and generates a light emission of the sputtered atoms. The specific emission lines of the single elements are optically diffracted and detected by several photomultipliers. For the quantification of the matrix elements, a reference measurement done by EDS is needed.

The $\Delta E_F$ measurements were performed by a home-built absolute PL set-up equipped with a 405 nm laser excitation source at room temperature using a CCD Si detector.[45] The experimental setup was first calibrated, and the resulting spectra were corrected spectrally and in intensity, as described in ref[66]. The incident photon flux was matched to a 1 sun spectrum above the bandgap of the absorber. The PL quantum yield ($Y_{PL}$) method was used to determine the $\Delta E_F$ using the PL maximum as the bandgap.[44,45]

The TRPL measurements were carried out using a time-correlated single photon counting (TCSPC) system with 638 nm wavelength pulsed laser at a repetition rate of 10 MHz, detecting the band edge emission with a bandwidth of 46 nm.

FIB was carried out using a FEI Helios Nanolab SEM/FIB using settings from reference[74]. For the preparation of the 20° bevel cross sections, the absorbers were tilted to 18° relative to the FIB ion milling gun. A 1 μm thick Pt layer was deposited on the absorber surface to minimize the curtaining effect. To reduce surface damage caused by the high energy Ga ion beam, the absorbers were further cleaned using the Gatan Ilion II broad ion beam (BIB) system. The absorber surface was milled by 0.5 kV broad Ar ion beam for 5 min at a temperature of 80 K. The BIB stage continuously rotated at 3 rotations per minute during the cleaning. Prior to EBSD and CL measurements, 5 nm of carbon was sputtered on the absorber surface to reduce charging.

The CL measurements were carried out by using an Attolight Allain 4027 Chronos dedicated CL-SEM at room temperature. The CL measurements on the bevel and cleaved cross sections were done with a 5 kV electron beam with a 100 μm and a 50 μm aperture, respectively. The plan view CL measurements were carried out with a 10 kV electron beam and a 50 μm aperture. The dwell time per pixel for all measurements was 250 ms. The CL data analysis was carried out using the open-source Python libraries, Hyperspy[75] and Lumispy[76].

The EBSD measurements were carried out on a Zeiss Gemini 300 SEM equipped with an Oxford Instruments HKL Symmetry 3 EBSD detector. All measurements used a 15 kV electron beam, a 50 ms pixel exposure time, and a 40 nm pixel step size. The EBSD indexing was carried out by Aztec HKL software, employing Hough-based indexing with a zinc blende cubic lattice (a = b = c = 5.58 Å). The EBSD analysis was carried out using the MATLAB toolbox MTEX[77].

Scanning transmission electron microscopy (STEM) images were acquired on a probe-corrected Thermo Fisher Spectra 300 S/TEM operated at 300 kV. The EDS signal was collected with a Dual-X system comprising two detectors, one on either side of the sample, for a total acquisition solid angle of 1.76 sr. Compositional maps were acquired using Velox, with repeated raster scanning and a probe current of ~150 pA. Elemental maps (Figure S9c) were produced with Velox. The GGI depth profile was calculated from the Ga and In signals after integration across the width of each map (~150 pixels). The 4-dimensional scanning transmission electron microscopy (4D-STEM) data was recorded on the same microscope, at 300 kV, with a camera length of 38 mm, step size of 2 nm and pixel time of 0.02 s. 4D-STEM data processing was carried out with two open-source Python libraries, pyxem[78] and py4DSTEM[37].

Solar cells were characterized using a class AAA solar simulator calibrated with a reference standard Si solar cell (RQN3154). The reference Si solar cell was calibrated in 2012. The devices were normally stored in the vacuum desiccator, except during the measurement and characterisation. The current-voltage (I-V) data was recorded at room temperature in the forward scan direction using an IV source measure unit with a scan speed of 50 mV/sec. The I-V measurements were performed under ambient conditions. The active solar cell area of ~0.42 cm$^2$ was determined using a Leica optical microscope and image processing software (ImageJ). The entire active area was illuminated with 1 sun illumination (100 mW/cm$^2$) during the measurement without any mask. The PCE of the solar cells was calculated based on the $J_{SC}$ obtained from the EQE integrated photocurrent density due to the higher spectral mismatch. EQE was measured using chopped illumination from a halogen-xenon lamp (Xenon short ARC lamp, Ushio uxl-302-0) and a lock-in amplifier to measure the photocurrent.

**Acknowledgements**

We would like to acknowledge funding from the Engineering and Physical Science Research Council (EPSRC) under EP/V029231/1 and EP/R025193/1. This work was also funded by Luxembourgish Fond National de la Recherche (FNR) for REACH (Project no: INTER/UKRI/20/15050982) and European Union within the SITA project (no. 101075626). We also acknowledge the support of the Wolfson Electron Microscopy Suite and use of the FEI Helios. Yucheng Hu would like to thank Ziyi Yuan and Xinjuan Li for support of the CL sample preparation.

**Author Contributions**

Conceptualization, S.S., G.K., and R.A.O.; Methodology, Y.H. and A.V.O.; Investigation – sample growth, A.V.O., M.M., K.K., A.H. and T.T.; Investigation – specimen preparation, Y.H., M.P. and S.M.; Investigation – GDOES, W.H. and W.W.; Investigation – PL, A.V.O.; Investigation – XRD, A.V.O. and M.M.; Investigation – SEM (EDS, CL, EBSD), Y.H., A.V.O., E.W. and M.P.; Investigation – STEM (EDS, 4D-STEM), A.G. and Y.P.I.; Investigation – Device Measurement, A.V.O. and M.M.; Visualization, Y.H, A.V.O., E.W. and A.G.; Writing – Original Draft Preparation, Y.H., A.V.O. and E.W.; Writing – Review & Editing, all; Supervision, G.D., S.S., G.K. and R.A.O.; Funding Acquisition, S.S., W.W., G.K. and R.A.O.

**Data Availability Statement**

All raw data generated in this study have been deposited to Cambridge Repository Apollo and can be openly accessed through following URL/DOI: https://doi.org/10.17863/CAM.118058.

# Supporting Information

**Silver alloyed wide bandgap (Ag,Cu)(In,Ga)S$_2$ thin film solar cell with 15.5% efficiency**

*Authors: Yucheng Hu\*, Ece Washbrook, Arivazhagan Valluvar Oli, Andrea Griesi, Yurii P. Ivanov, Mariam Pelling, Simon M. Fairclough, Kulwinder Kaur, Michele Melchiorre, Adam Hultqvist, Tobias Törndahl, Wolfram Hempel, Wolfram Witte, Giorgio Divitini, Susanne Siebentritt, Rachel A. Oliver, Gunnar Kusch\**

*Y. Hu, E. Washbrook, and A. V. Oli contributed equally to this work.*

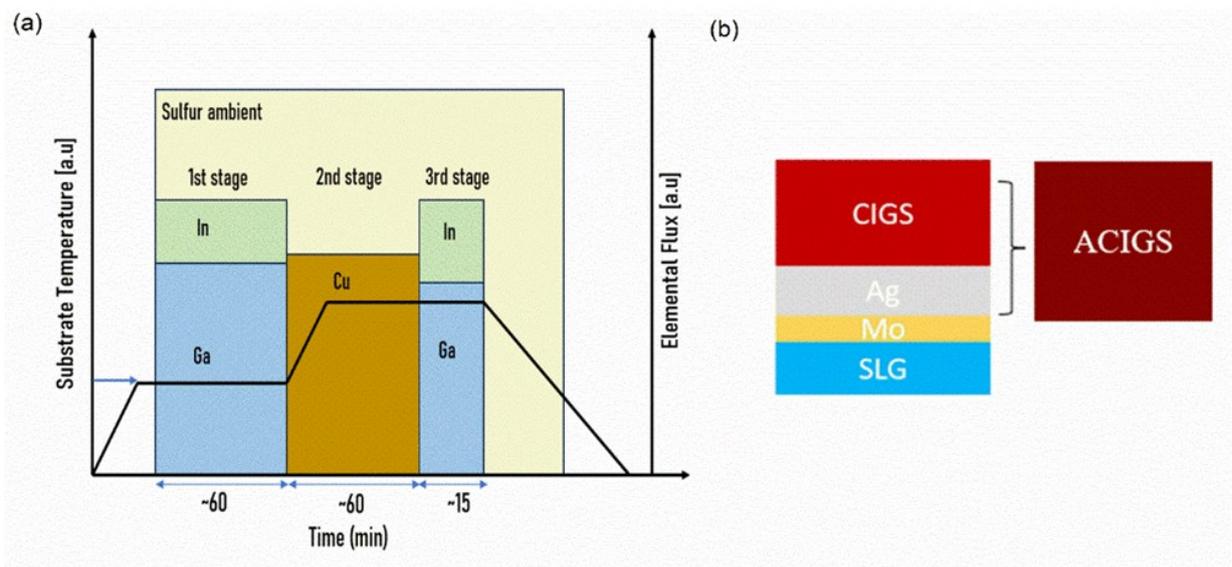

*Figure S1:* *(a) Deposition profile of CIGS. The bars in the deposition profiles are the corresponding elemental fluxes and the black line is the substrate temperature. The first stage is at 470 °C, and the second and the third stages are at 600 °C. (b) Conversion scheme illustrating transformation of Ag precursor and CIGS deposited at elevated temperature into ACIGS.*

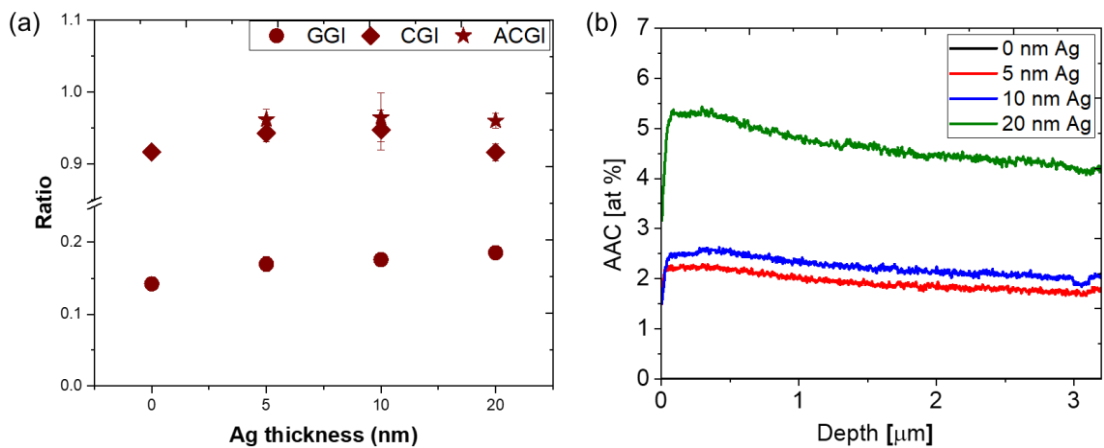

*Figure S2:* *Surface GGI, CGI and ACGI ratios with varying Ag thicknesses, performed at 7 kV, by EDS. (b) GDOES AAC depth profiles of ACIGS from top (left) to bottom (right). Related to main text* ***Figure 1.***

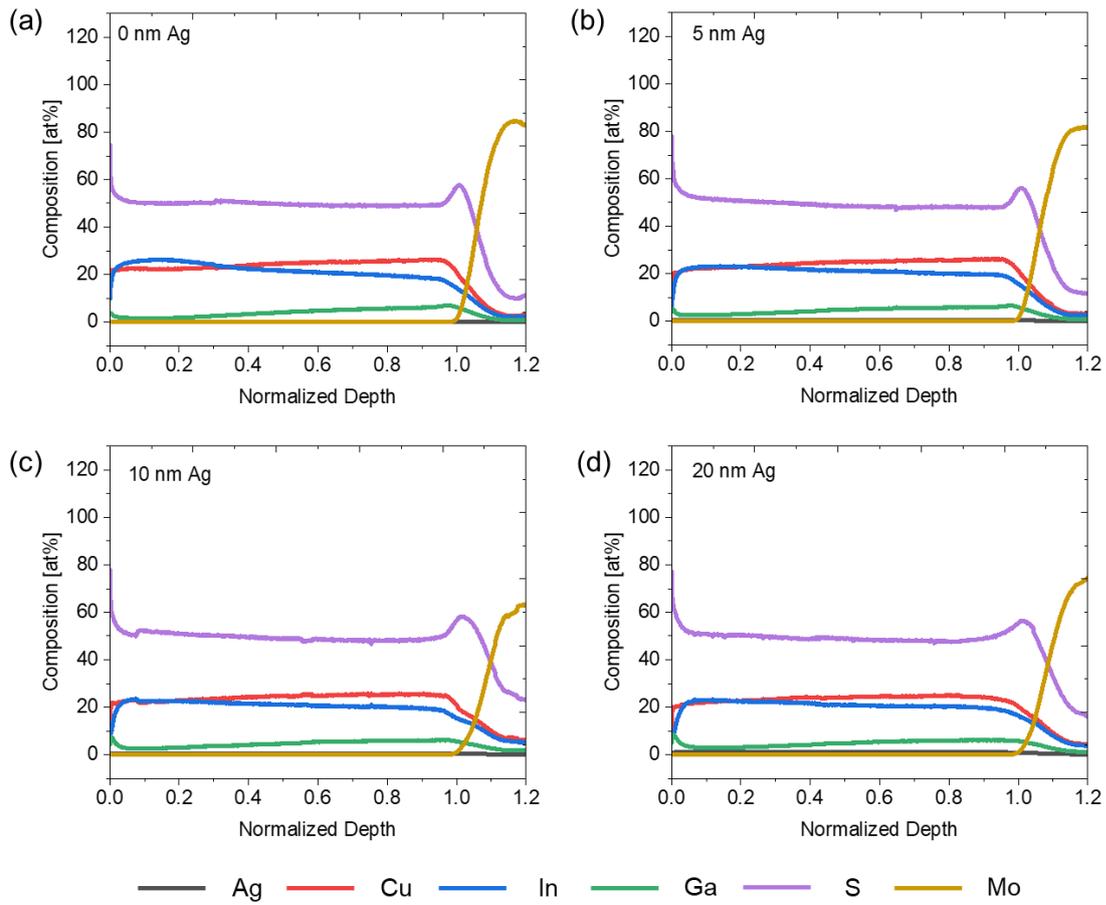

*Figure S3:* Quantified GDOES depth profiles of various elements in (A)CIGS with different Ag precursor layer thicknesses. Related to main text *Figure 1*.

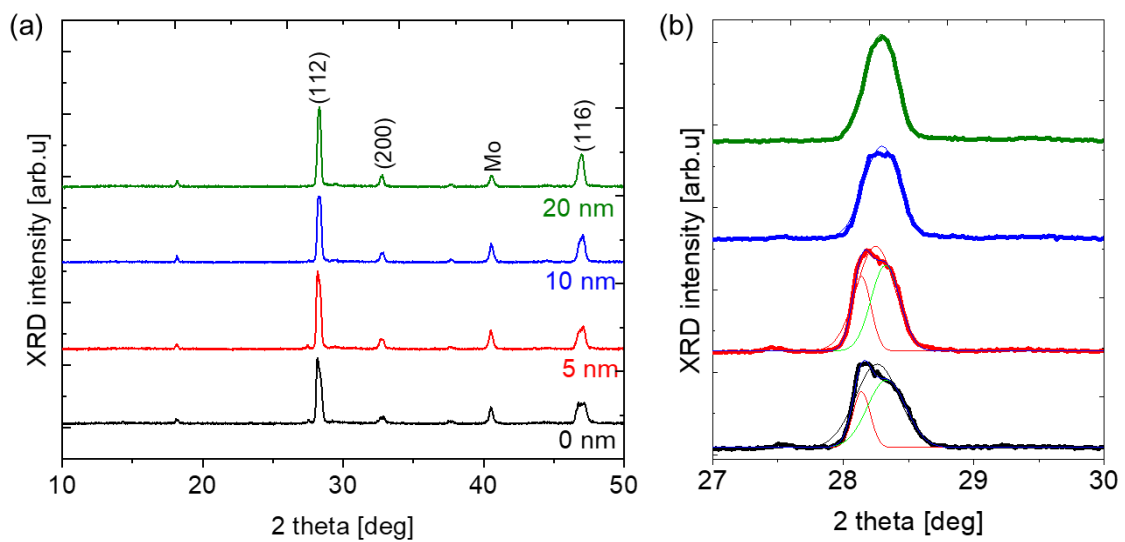

*Figure S4:* (a) Wide angle XRD pattern of ACIGS with different Ag thicknesses; (b) Gaussian fit of the 112 reflex of the XRD patten. Related to main text *Figure 2.*

*Table S1.* Proportion of voids area occupied on the bevel cross sectional surface. The bevels are equally divided into three sections to count the void density variation along the depth changes.

|  |  | **0 nm Ag** | **5 nm Ag** | **10 nm Ag** | **20 nm Ag** |
|---|---|---|---|---|---|
| Overall |  | 13.5% | 9.0% | 7.0% | 5.0% |
|  | Front | 15.0% | 7.0% | 2.2% | 0.4% |
|  | Middle | 13.0% | 8.0% | 9.0% | 4.4% |
|  | Back | 13.5% | 8.5% | 12.5% | 7.5% |

*Table S2.* Grain size and GBs length quantified from EBSD maps. Similar to the void density estimate, the bevels were split into three sections to calculate the microstructural variation along the depth of the absorber.

|  |  | **0 nm Ag** | **5 nm Ag** | **10 nm Ag** | **20 nm Ag** |
|---|---|---|---|---|---|
| Average Grain Size [$\mu m^2$] |  | 0.164 | 0.266 | 0.526 | 1.072 |
|  | Front | 0.214 | 0.299 | 0.596 | 1.020 |
|  | Middle | 0.135 | 0.217 | 0.510 | 1.664 |
|  | Back | 0.105 | 0.201 | 0.329 | 0.694 |
| Total GB [$\mu m$] |  | 565.3 | 373.5 | 234.0 | 209.7 |
|  | Front | 198.0 | 150.8 | 101.0 | 82.5 |
|  | Middle | 232.9 | 163.3 | 94.6 | 49.5 |
|  | Back | 272.9 | 171.3 | 119.7 | 79.4 |

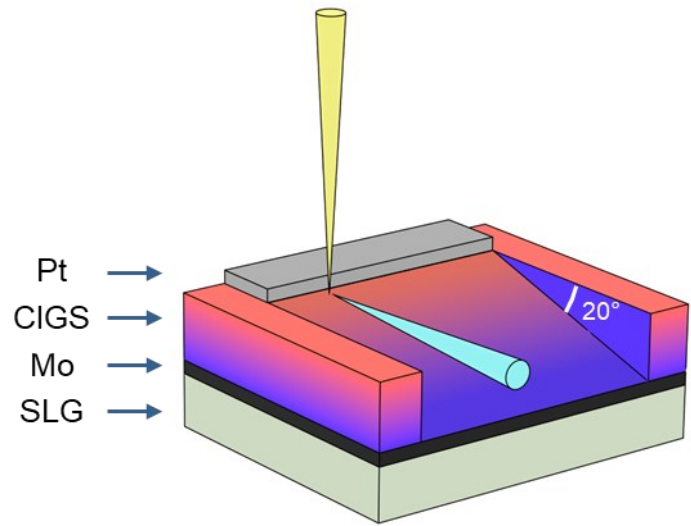

***Figure S6:*** *Schematic shows the beveled cross section for bevel SEM (Figure 3a), EBSD (Figure 3b) and CL (Figure 5) measurements. Yellow cone highlights the electron beam incident direction for bevel SEM and CL measurements, while blue cone presents the electron beam direction for EBSD. Because electron beam for bevel SEM and CL incidence in the direction perpendicular to the as-grown surface plane, instead of perpendicular to the bevel surface plane, the length of bevel SEM images and CL maps acquired should be slightly shorter than the length of actual bevel surface. The slight difference in projection should make no difference to the void counting and opto-electronic properties extracted from bevel SEM and CL.*

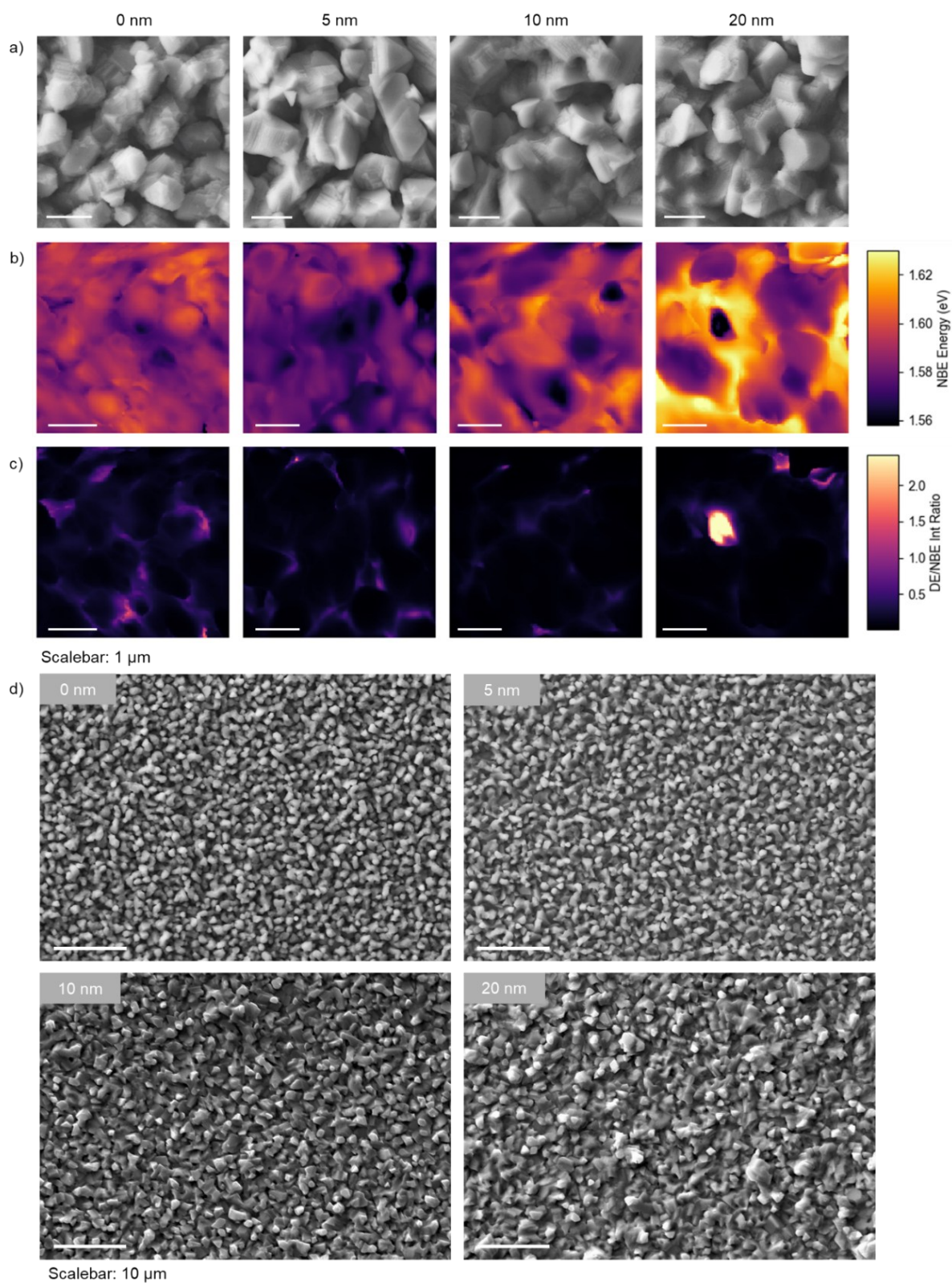

*Figure S5:* (a) SEM images of plan-view CL mapped area, with corresponding b) NBE peak energy distribution map and c) DE/NBE intensity ratio map; d) large area SEM image of absorbers with different Ag contents.

*Table S3. Tri-exponential fit values of TRPL transients and calculated doping density. Related to Fig. 3c and 3d.*

| Ag thickness [nm] | $A_1$ | $A_2$ | $A_3$ | $\tau_1$ [ns] | $\tau_2$ [ns] | $\tau_3$ [ns] | $\tau_{eff}$ [ns] | $N_A$ [×17 cm$^{-3}$] |
|---|---|---|---|---|---|---|---|---|
| 0 | 0.1 | 0.1 | 0.01 | 1.2 | 3.8 | 14.8 | 3.2 | 1.09 |
| 5 | 0.2 | 0.2 | 0.07 | 1.3 | 5.1 | 22..2 | 5.6 | 1.36 |
| 10 | 0.3 | 0.3 | 0.07 | 1.8 | 6.5 | 28.5 | 7.5 | 1.01 |
| 20 | 0.2 | 0.3 | 0.1 | 1.8 | 7.3 | 27.6 | 10.6 | 1.05 |

*Table S4. Fitted parameters of plan view CL spectra NBE peak.*

|  | 0 nm Ag | 5 nm Ag | 10 nm Ag | 20 nm Ag |
|---|---|---|---|---|
| FWHM (eV) | 0.094 | 0.095 | 0.096 | 0.103 |
| Intensity (arb.u.) | 438.8 | 1113.2 | 1677.7 | 1812.3 |
| Energy (eV) | 1.588 | 1.577 | 1.583 | 1.594 |

*Table S5. NBE peak energy standard deviation for different parts of bevel cross section.*

|  |  | 0 nm Ag | 5 nm Ag | 10 nm Ag | 20 nm Ag |
|---|---|---|---|---|---|
| Mean NBE Energy std (meV) |  | 81 | 80 | 51 | 52 |
|  | Front | 29 | 33 | 32 | 35 |
|  | Middle | 38 | 33 | 25 | 26 |
|  | Back | 30 | 34 | 17 | 21 |

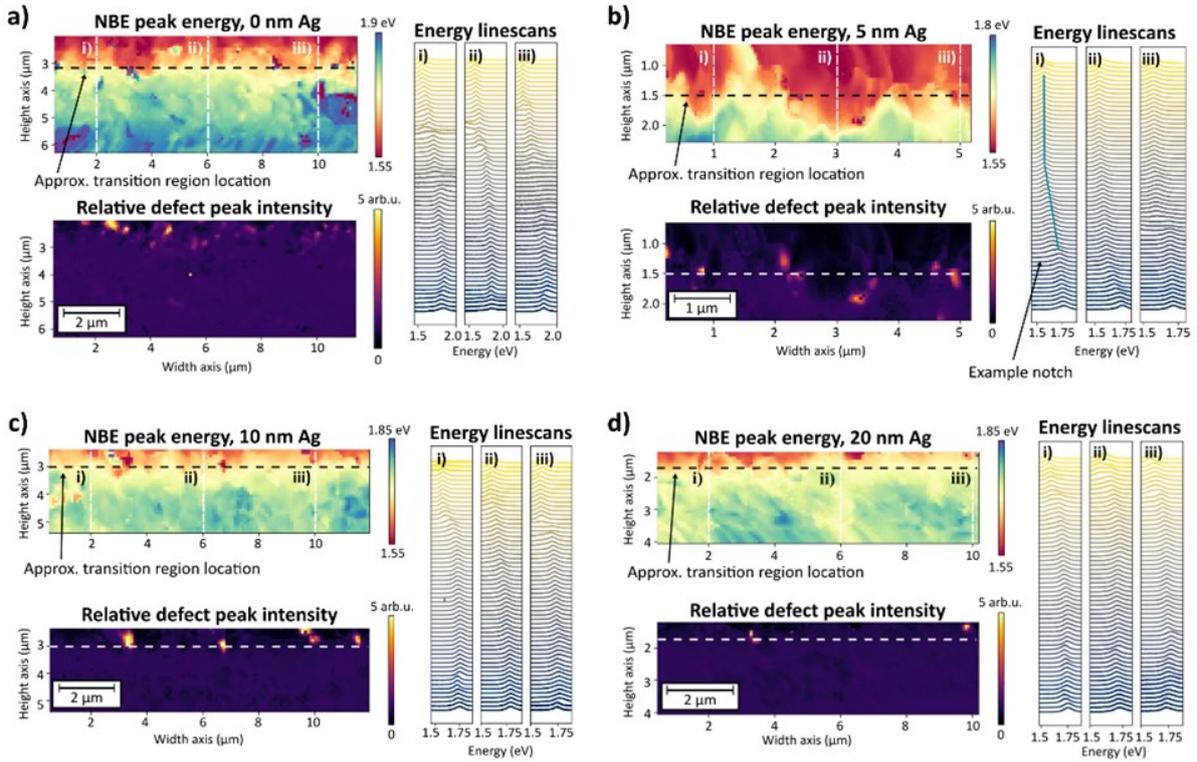

*Figure S7:* The CL results on cleaving cross sections of ACIGS absorbers with different Ag precursor thicknesses, including the NBE emission energy map, relative defect peak intensity (DE/NBE) and three linescans extracted at position i), ii), and iii). Strong DE spots can be found in all four samples.

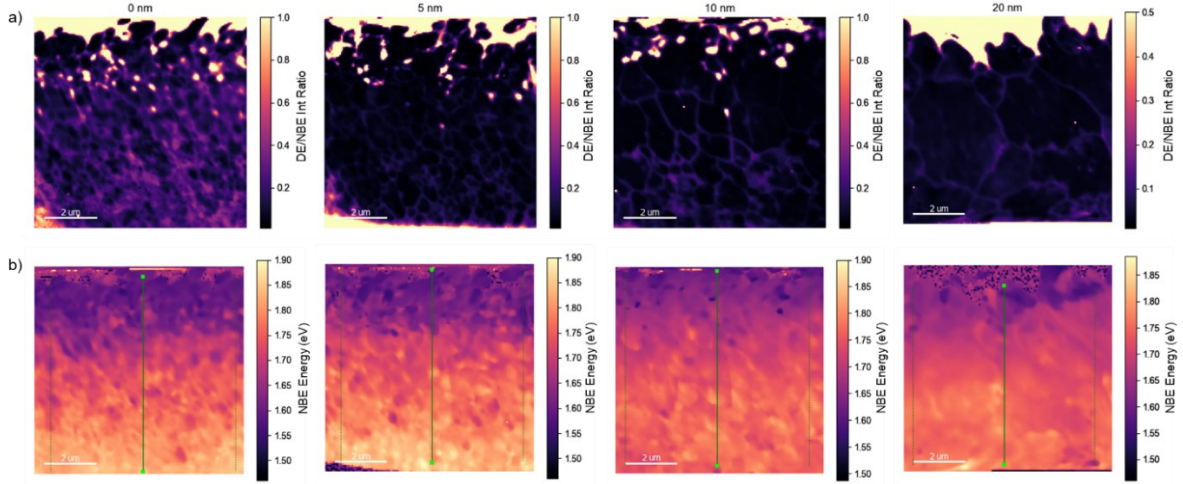

*Figure S8:* a) DE/NBE Int Ratio maps for four bevel cross section with modified maximum value to highlight the presence of RHAGBs, the DE is comparably dim at RHAGB in 20 nm; b) Line markers exhibited the extracted positions for line profile and line spectraces shown in Fig 5. All linescans have been integrated across a width of 7.3 μm, to mitigate the influence of GBs on linescans.

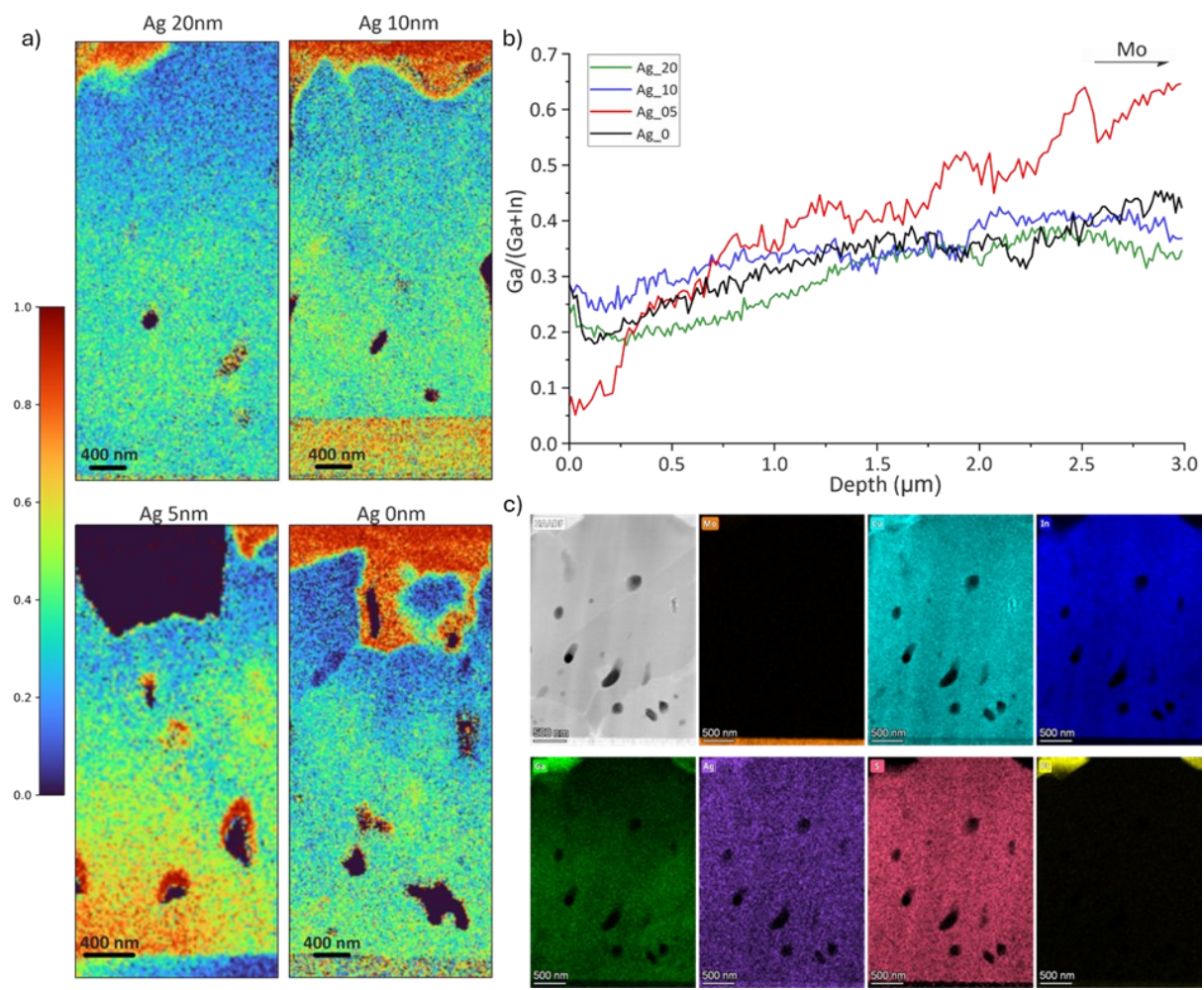

*Figure S9:* STEM-EDS maps on CIGS and ACIGS absorbers. a) Variation of GGI ratio across the map; b) GGI profile along the depth, integrated from the maps in a) - the profile can be noisy due to the presence of surface morphology and voids; c) elemental distribution maps of Mo (back contact), Cu, In, Ga, Ag, S, and Pt (FIB capping layer) on 20 nm ACIGS.

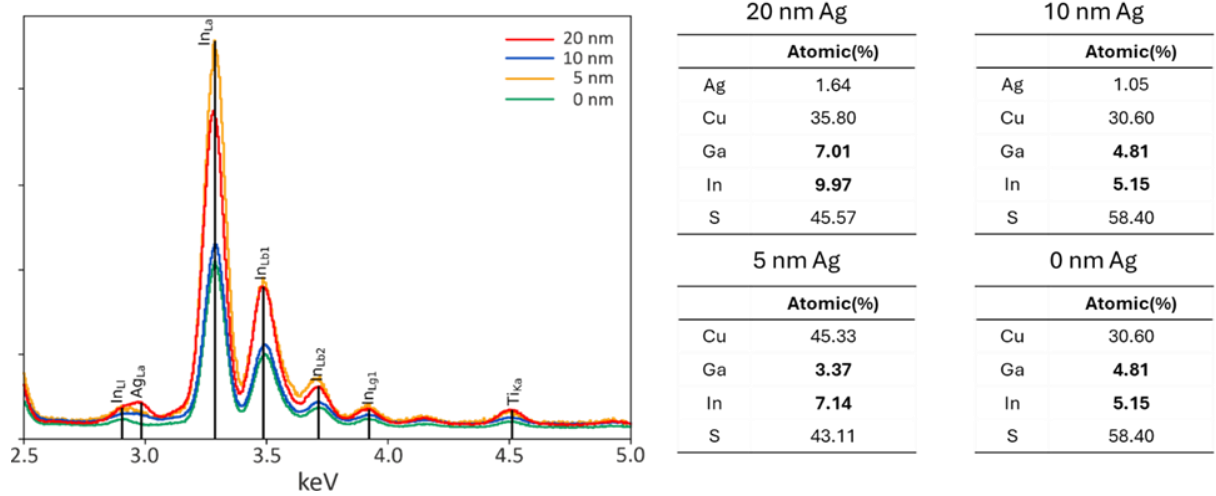

*Figure S10:* STEM-EDS spectra for all four samples used for EBSD and CL measurements. The extracted atomic percentage of is shown in table aside. The Ag content in 5 nm Ag is very low and inhomogeneous.

*Table S6.* $E_g$ obtained from PL, CL, and EQE derivative.

|  | 0 nm Ag | 5 nm Ag | 10 nm Ag | 20 nm Ag |
| --- | --- | --- | --- | --- |
| $E_g(PL)$ | 1.55 | 1.56 | 1.56 | 1.56 |
| $E_g(CL)$ | 1.59 | 1.58 | 1.58 | 1.59 |
| $E_g(EQE)$ | 1.56 | 1.58 | 1.58 | 1.58 |

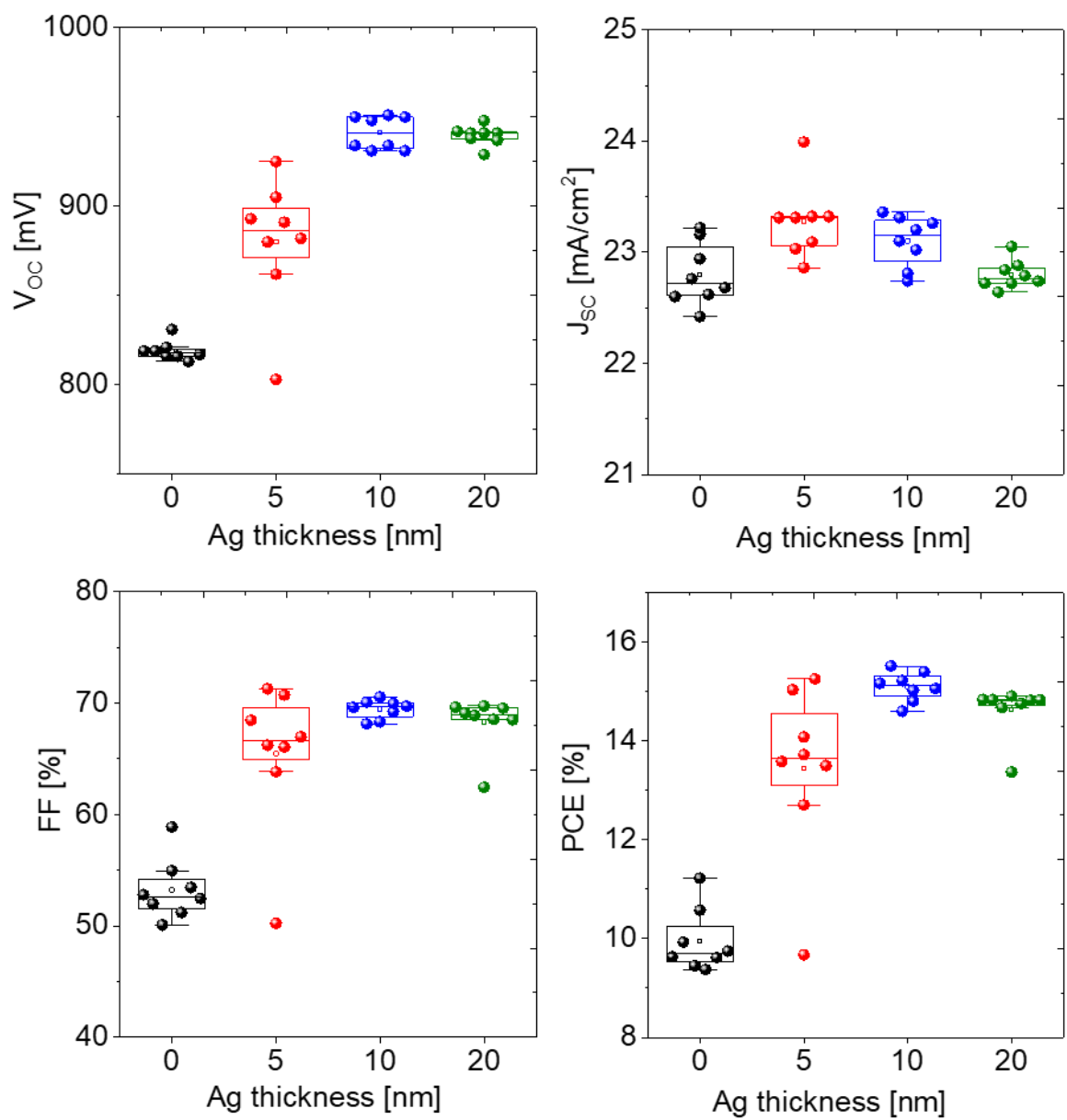

**Figure S11:** *Statistical distribution of the photovoltaic parameters for solar cells with varying Ag thicknesses.*

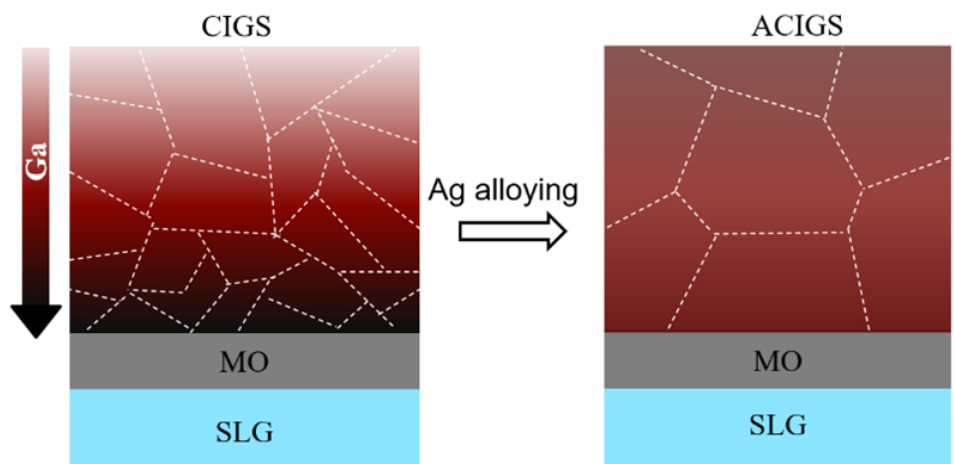

*Figure S12:* *Schematic representation of the mechanism underlying Ag alloying in CIGS, illustrating the formation of the larger grains and improved distribution of Ga throughout the depth of the absorber.*

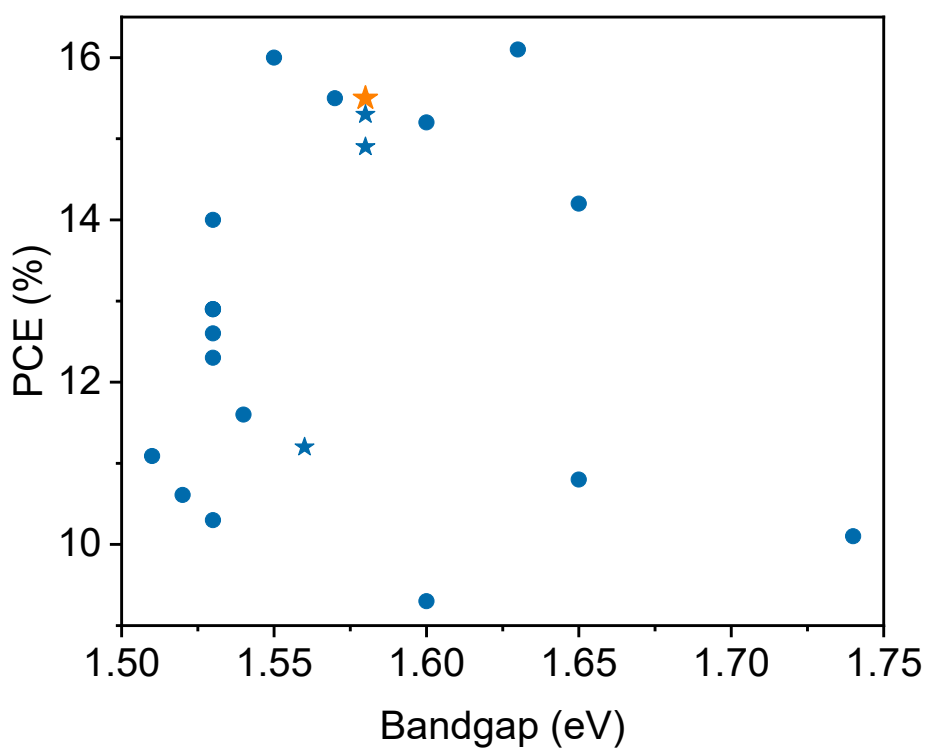

*Figure S13:* *Comparison of power conversion efficiency (PCE) between our solar cells and reports in CIGS reference[1–12]. The star marked points are solar cells reported in this manuscript and the orange point marked our champion device. Bandgap here refers to bandgap values derived from EQE curves.*

**Supplementary Note 1: Comparison between bevel and polished cross section characterization results**

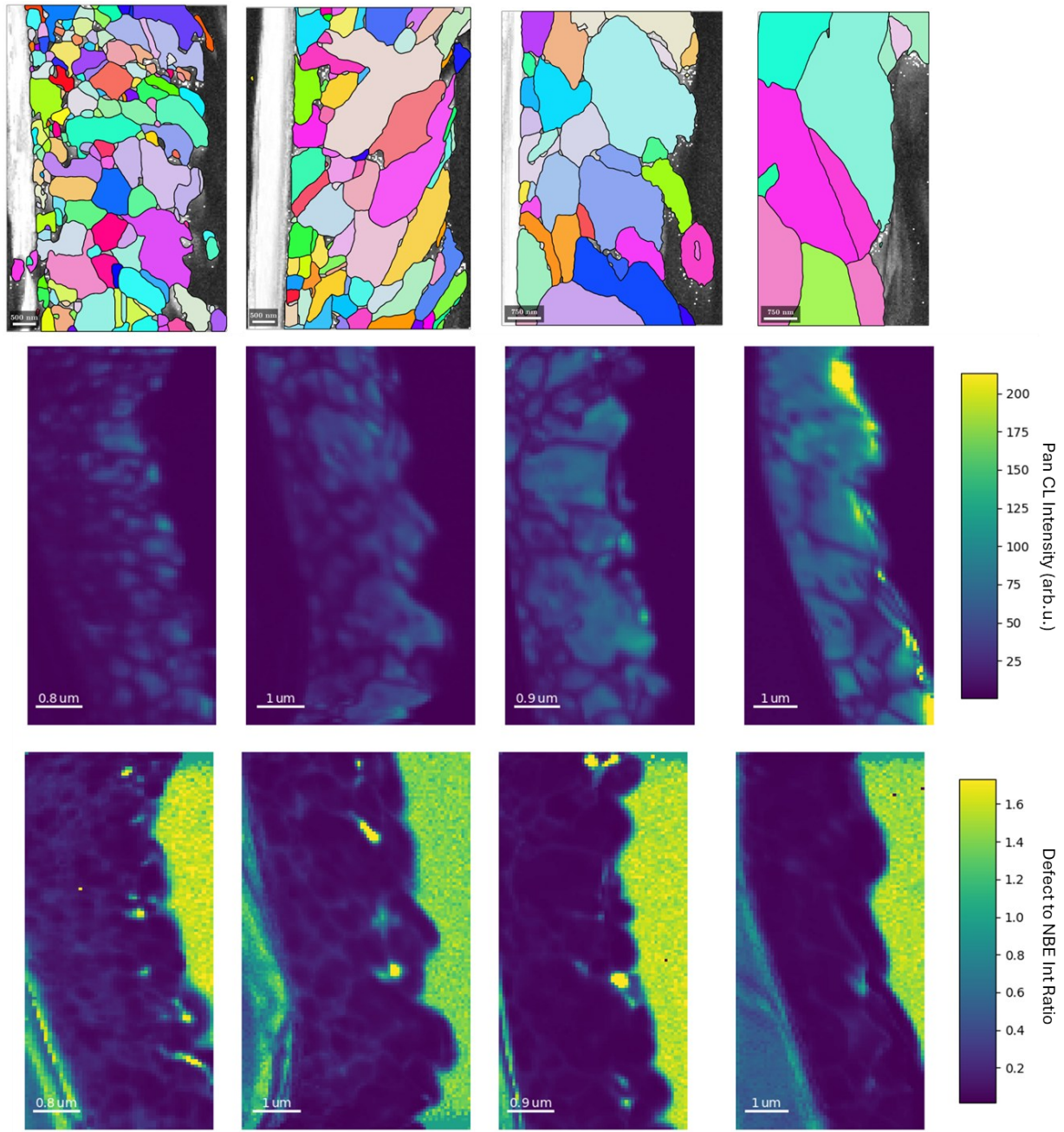

*Figure S14: Correlative EBSD (top row) CL (middle and bottom rows) results on polished cross sections, present in CIGS, 5 nm, 10 nm, and 20 nm ACIGS order (from left to right).* In order to achieve proper characterization of absorber cross sectional information with minimal influence of cleaving morphology, three sample preparation strategies have been tested, including the polished cross section, the bevel method, and the FIB lift-out lamella. The Figure S14 shows the correlative EBSD CL results of polished cross section, which exhibit map drift due to strong charging effect caused by soda-lime glass substrate. The polished cross section was prepared according to the concept given by D. Abou-Ras et al.,

employing mechanical polishing followed by broad ion beam polishing[13]. With this method, we achieved sensible EBSD and CL results with clear grain structure. In terms of microstructural information, such as grain size variation, the polished cross section shows the same trend as EBSD results on bevel structure present in main text. However, the grain structure, especially the 5 nm and 20 nm samples, exhibited definite oriented shape, which can be related to sample distortion caused by mechanical polishing. The difference in EBSD quality between our polished cross section and literature may be due to the difference in sample softness and/or mechanical polishing parameters. The grain distortion not only can influence the quantitative analysis of microstructure, but also adding complicity to the correlation between EBSD and CL maps, especially for highly polycrystalline samples.

FIB lift-out lamella is another potential strategy for cross-sectional characterization. Nevertheless, the high energy Ga ion beam damage can have strongly influence CL measurements by introducing high density of surface defects, leading to weak or even no contrast between grain interior and GBs. In bevel method, FIB was also used to make such an inclined surface on sample. To minimize the influence of FIB damage, the bevel surface was then cleaned using low energy Ar broad ion beam polishing.

With considerations of both results quality and reproducibility of sample preparation, we decided to use bevel method for cross sectional characterization in this manuscript.

**Reference for SI**